\documentclass[10pt]{IEEEtran}

\usepackage{booktabs}
\usepackage{url}
\usepackage[pdftex]{graphicx}
\usepackage{float}
\usepackage{epstopdf}
\usepackage{amsmath}
\usepackage{amssymb}
\usepackage{pifont}
\usepackage{color}
\usepackage[table,xcdraw]{xcolor}
\usepackage{array}
\usepackage{verse}
\usepackage{cite}

\begin{document}
\title{\textcolor{black}{The Past, Present, and Future of Transport-Layer Multipath}}
\author{Sana Habib, Junaid Qadir, Anwaar Ali, Durdana Habib, Ming Li, Arjuna Sathiaseelan 

\thanks{Sana Habib (14mseeshabib@seecs.edu.pk) is a postgraduate student of Electrical Engineering at the Electrical Engineering Department of SEECS, NUST. \textcolor{black}{Junaid Qadir (junaid.qadir@itu.edu.pk) is an Associate Professor at the Information Technology University-Punjab, Lahore, Pakistan.} Anwaar Ali (anwaar.ali@itu.edu.pk) is a Research Associate at the Information Technology University-Punjab, Lahore, Pakistan. \textcolor{black}{Durdana Habib (durdana.habib@nu.edu.pk) is an Assistant Professor at the National University of Computer and Emerging Sciences, Islamabad, Pakistan.} Ming Li (ming.li@aalto.fi) is with the Data
Communications Software Research Group from the Dept. of Computer
Science, Aalto Unviersity, Finland. Arjuna Sathiaseelan (arjuna.sathiaseelan@cl.cam.ac.uk) is a Senior Research Associate at the Computer Laboratory, University of Cambridge. 
}
} 
\maketitle

\begin{abstract}
Multipathing in communication networks is  gaining momentum due to its attractive features of increased reliability, throughput, fault tolerance\textcolor{black}{,} and load balancing capabilities. In particular, \textcolor{black}{wireless environments and datacenters are envisioned to become} largely dependent on the power of multipathing for seamless handovers, virtual machine (VM) migration and in general, pooling less proficient resources together for achieving overall high proficiency. The transport layer, \textcolor{black}{with its} knowledge about end-to-end path characteristics, is well placed to enhance performance through better utilization of multiple paths. \textcolor{black}{Realizing the importance of transport-layer multipath, this paper investigates the modernization of traditional \textcolor{black}{connection establishment,} flow control, sequence number splitting, acknowledgement\textcolor{black}{,} and flow scheduling mechanisms for use with multiple paths.} 
\textcolor{black}{Since congestion control defines fundamental feature of the transport layer, we study the working of multipath rate control and analyze its stability and convergence. We also discuss how various multipath congestion control algorithms differ in their window increase and decrease functions, their TCP-friendliness, and responsiveness. 
To the best of our knowledge, this is the first in-depth survey paper that has chronicled the evolution of the transport layer of the Internet from the traditional single-path TCP to the recent development of the modern multipath TCP (MPTCP) protocol. Along with describing the history of this evolution, we also highlight in this paper the remaining challenges and research issues.}
\end{abstract}


\section{Introduction}
\begin{quote}
\textcolor{black}{\textit{``The best way to predict the future is to create it."--- Abraham Lincoln}}
\end{quote}

Traditionally, transport layer uses a single path for end-to-end communication between applications. However, single-path transport protocols are not able to keep up with the growing \textcolor{black}{bandwidth as well as}  reliability and fault tolerance demands of multimedia applications and critical businesses (\textcolor{black}{such as communication between government agencies and \textcolor{black}{e-commerce}}). This deficiency of \textcolor{black}{the} single-path protocols has led to the increasing trend of multipathing in the Internet. 

Multipathing elegantly solves the deficiencies of single-path protocols by employing \emph{inverse multiplexing} of resources to send packets over a set of paths rather than a single-path. This striping of data across multiple paths provides better reliability, fault tolerance\textcolor{black}{,} and increased throughput \cite{ye2008improving,kim2005improving,fastestTCP}. Thus, applications can reap the benefits of resource pooling and diversity provided by multiple paths to achieve desired quality of service (QoS) as well as improved user's quality of experience (QoE). \textcolor{black}{Instead of discussing multipathing at different protocol stack layers in terms of performance enhancements \cite{qadir2015exploiting,singhsurvey}; this paper focuses on the use of multiple paths at the transport layer.}

Multipath transport layer is well suited to devices equipped with \textcolor{black}{heterogeneous access technologies} (such as a mobile device equipped with both Wi-Fi and 3G) \cite{paasch2012exploring}. Use of multiple paths by mobile devices not only improves reliability  by providing the ability to seamlessly shift from one technology to another (that can act as the backup interface in the case of outage or congestion) but users can also benefit from cheap prices (e.g., when both 3G and Wi-Fi are available, the latter can be used being \textcolor{black}{less expensive}). \textcolor{black}{It can also become possible for mobile nodes in close proximity to pool their low bandwidths in order to support high bandwidth applications \cite{kim2005improving}.}
Recent implementation of MPTCP by Apple iOS7 \cite{iOS} has further encouraged the multipath trend at the transport layer. 

Datacenters in particular, represent an important use case of transport-layer multipathing. \textcolor{black}{\textcolor{black}{Also, multipath} transport layer enables many topologies in datacenter that could not be realized with single-path TCP. For example, GRIN \cite{agache2012grin} uses MPTCP to efficiently utilize datacenter network by making minor topology changes. Raiciu et al. \cite{raiciu2011improving} showed that Amazon EC2 achieves three times throughput in comparison to single-path by exploiting path diversity.}

\textcolor{black}{To further emphasize on the importance of multipath at transport layer, we next describe some of the benefits in detail.}

\subsection{Merits of Multipathing}
\label{Benefits of Multipathing}
Four major benefits obtained by using multiple paths are mentioned below.

\subsubsection{Load Balancing}
\textcolor{black}{Load balancing \textcolor{black}{is} traditionally performed at the network layer, wherein a network operator \textcolor{black}{can} reroute traffic in order to avoid congested hotspots. However, traffic engineering at the network layer can be unstable and may lead to oscillations \cite{abt2009trilogy}.} \textcolor{black}{Let us elaborate this argument with an example. Consider \textcolor{black}{that} a source-destination pair is using a path\textcolor{black}{,} say $A$\textcolor{black}{,} for communication and it is initially congested. Later, a better path $B$ becomes available. Now, it is desirable to balance traffic between paths $A$ and $B$. 
It is possible for the network layer to naively shift the bulk of the traffic to path $B$, thereby moving the congestion to path $B$ rather than load balancing. With path $B$ now becoming congested, the network layer will then force the traffic back to path $A$, leading to oscillations and instability. Transport layer on the other hand, gradually increases congestion window size on path $B$ over a period of several round-trip time (RTTs) and balances traffic between $A$ and $B$ in a more stable fashion.}

\textcolor{black}{Thus,} use of multiple paths at the transport layer can enable better load balancing properties due to the availability of \textcolor{black}{fine-grained} information about \textcolor{black}{the network's} end-to-end characteristics \textcolor{black}{such as available bandwidth, RTT, etc}. \textcolor{black}{This compelling observation was made by} the research of Kelly and Voice \cite{kelly2005stability}. In particular, the multipath transport protocols are designed to move traffic from more congested to less congested paths. With this shifting, the loss rates on less congested path increase and on more congested path decrease; the overall result is that the loss rates across the network tend to equalize \cite{dong2007multi}. 

 

\subsubsection{Resource Pooling} 
The idea of pooling resources together into one single resource, which has the aggregate abilities of all the resources, has been widely used in the Internet \cite{wischik2008resource}. Specifically, pooling of multiple paths at the transport layer provides better aggregate path characteristics (\textcolor{black}{e.g.,} bandwidth, delay\textcolor{black}, and RTT) \textcolor{black}{than each individual paths.}  Resource pooling allows to efficiently use network resources by dynamically allocating resources to meet the traffic surges. Once the traffic transaction is complete, the resources go back in the pool. 
Resource pooling enables the Internet to have higher reliability and robustness than the individual links and routers. 

In single-path applications, if a path fails then alternate paths can be used. However, the failure of primary path causes a temporary interruption in application until an alternate path is established (e.g., transient problems on a radio interface).  Pooling of multiple paths on the other hand, enables the network to transparently shift traffic from faulty paths to non-faulty paths without any interruption in the operation of an application. 

\subsubsection{Diversification}
Diversity is a technique that is frequently deployed in datacenters, wireless environments\textcolor{black}{,} and the Internet to improve performance. It has been shown in literature that the default (single) path between source and destination is often not the best path. A measurement study \cite{savage1999end} showed that in 30-80\% of the cases, an alternate path with better quality in comparison to the default path is available. By exploiting the diversity in the characteristics of multiple paths, performance enhancements including \textcolor{black}{bandwidth aggregation and reliability can be achieved}. Flow splitting at the transport layer by taking into account the diverse path characteristics (e.g., size of congestion window, RTT\textcolor{black}{,} and bandwidth) can further lead to better traffic engineering capabilities. 

In case of multimedia applications, streaming data over multiple paths can achieve better throughput and error resilience in comparison to \textcolor{black}{single path} by exploiting the heterogeneous path characteristics. The diversity of multiple paths has proved to be helpful in overcoming loss \cite{apostolopoulos2000reliable} and delay \cite{liang2001real} problems in multimedia applications.

\subsubsection{Role in the Future Internet Architectures}
\label{Role in the Future Internet Architectures}
The use of multiple paths have already enabled transparent hand overs between \textcolor{black}{heterogeneous access technologies} (e.g., between 3G and Wi-Fi) \textcolor{black}{\cite{paasch2012exploring}}, seamless virtual machine migration for power efficiency\textcolor{black}{,} and are envisioned to play a key role in the future Internet architectures like 5G and cloud. In particular, 5G is envisioned to use concurrent multiple paths for data transfer in order to meet the high bandwidth requirements while providing robustness and reliability \cite{hossain20135g}. Moreover, the increased throughput and connection resiliency requirements of clouds can also be met with the use of multiple paths.
\\

\begin{table*}[!ht]
\scriptsize
\centering
\caption{The Coverage of Topics in This and Related Surveys}
\label{tab:Related Surveys}
\begin{tabular}{p{2cm}p{0.4cm}p{1cm}p{1cm}p{1cm}p{1cm}p{1.25cm}p{1.2cm}p{1cm}p{1.85cm}p{1.6cm}}

\hline
\cellcolor[HTML]{EFEFEF}\textbf{\emph{Survey}} 
& \cellcolor[HTML]{EFEFEF}\textbf{\emph{Year}}
&


\cellcolor[HTML]{EFEFEF}\textbf{\textcolor{black}{\emph{Connection setup}}}
& \cellcolor[HTML]{EFEFEF}\textbf{\emph{Flow Control}}
& \cellcolor[HTML]{EFEFEF}\textbf{\emph{Sequence Number Splitting}}
& \cellcolor[HTML]{EFEFEF}\textbf{\emph{ACK}}
&
\cellcolor[HTML]{EFEFEF}\textbf{\emph{Flow Scheduling}}
&
\cellcolor[HTML]{EFEFEF}\textbf{\emph{NUM Framework}}
&
\cellcolor[HTML]{EFEFEF}\textbf{\textcolor{black}{\emph{Congestion Control}}}
&
\cellcolor[HTML]{EFEFEF}\textbf{\emph{\textcolor{black}{Window-Increase/Window-Decrease} function}}
&
\cellcolor[HTML]{EFEFEF}\textbf{\emph{TCP-Friendliness \& Responsiveness}}
\\
\hline

Zhuang et al. \cite{zhuang2012multipath} & 2012 & \textcolor{black}{$\times$} & $\checkmark$ & $\checkmark$ & $\checkmark$ & $\times$ & $\times$ & \textcolor{black}{$\times$} & $\times$ & $\times$ \\
Ramaboli et al. \cite{ramaboli2012bandwidth} & 2012 & \textcolor{black}{$\times$} & $\checkmark$ & $\checkmark$ & $\checkmark$ & $\times$ & $\times$ & \textcolor{black}{$\checkmark$} & $\times$ & $\times$\\
Habak et al. \cite{habak2013bandwidth} & 2013 & \textcolor{black}{$\times$} & $\checkmark$ & $\times$ & $\checkmark$ & $\checkmark$ & $\times$ & \textcolor{black}{$\checkmark$} & $\times$ & $\times$\\
Addepalli et al. \cite{addepalli2013heterogeneous} & 2013 & \textcolor{black}{$\times$} & $\checkmark$ & $\times$ & $\times$ & $\times$ & $\times$ & \textcolor{black}{$\checkmark$} & $\times$ & $\times$ \\

Singh et al. \cite{singhsurvey} & 2015 & \textcolor{black}{$\times$} & $\times$ & $\times$ & $\times$ & $\times$ & $\checkmark$ & \textcolor{black}{$\checkmark$} & $\times$ & $\times$ \\
Qadir et al. \cite{qadir2015exploiting} & 2015 & \textcolor{black}{$\times$} & $\times$ & $\times$ & $\times$ & $\times$ & $\times$ & \textcolor{black}{$\checkmark$} & $\times$  & $\times$  \\
This survey & 2015 & \textcolor{black}{$\checkmark$} & $\checkmark$ & $\checkmark$ & $\checkmark$ & $\checkmark$ & $\checkmark$ & \textcolor{black}{$\checkmark$} & $\checkmark$ & $\checkmark$ \\

\hline
\end{tabular}
\end{table*}

Having established the importance of transport-layer multipath in current and future Internet architectures, we next describe the design challenges that need to be addressed for successful deployment of multipath transport layer. 

\subsection{Challenges in Implementing Multipath Transport Layer}
\label{Challenges in Implementing Multipath Transport Layer}
Generally, there are three basic challenges that must be addressed by the transport layer so that multipath transmissions can work efficiently in heterogeneous environments (wired and wireless) that characterize the Internet.
 
The \emph{first challenge} with the use of multiple paths is the \textcolor{black}{increased packet reordering introduced at the receiver.} This is due to the fact that packets travel along paths \textcolor{black}{with varying delay characteristics.} \textcolor{black}{If the diverse nature of
different paths are not considered}, then unnecessary retransmissions may occur that are not due to lost but delayed packets. These  unnecessary retransmissions not only waste bandwidth but also violate the goal of minimizing congestion in the network. 

The \emph{second challenge} is the handling of data streams in concurrent transmissions. The data transmitted through multiple paths has to be reconstructed at the receiver. Thus, use of sequence numbers for loss detection and reconstruction of data transmitted through multiple paths is a critical design concern. The research community has also realized the importance of middleboxes in this regard.  Recent studies have shown that middleboxes do not simply forward packets but also modify packet headers that further complicates the problem \textcolor{black}{\cite{honda2011still, sherry2012making,hesmans2013tcp}.} 

\begin{figure} 
\centering
\includegraphics[width=8cm]{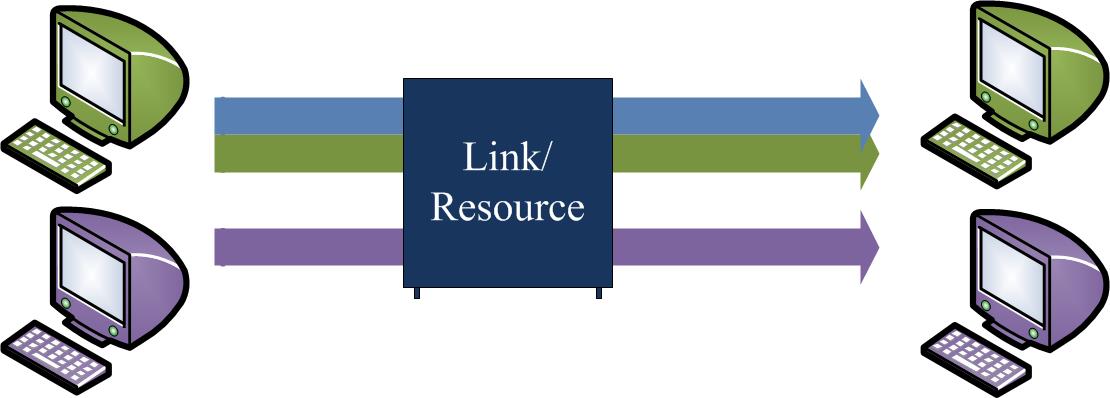}
\caption{Fairness issue with multipath transport layer.}
\label{fig1}
\end{figure}

The \textcolor{black}{\emph{third challenge} is to ensure that multipath traffic is fair to other traffic} (typically TCP flows, as they comprise 80-90\% of the total Internet traffic). In order to understand the problem, consider Fig. \ref{fig1} which shows two communicating pairs: 
one with multihoming capability---i.e., there are multiple paths between source and destination--- while the other only has a single-path connecting the source and destination. If there is no congestion at the link traversed by the two flows then there is no issue regarding fairness. However, in the case of a bottleneck link, the multihomed source-destination pair is unfair to the single-path pair as it receives twice the bandwidth. 

\begin{figure*}
\centering
\includegraphics[width=0.76 \textwidth]{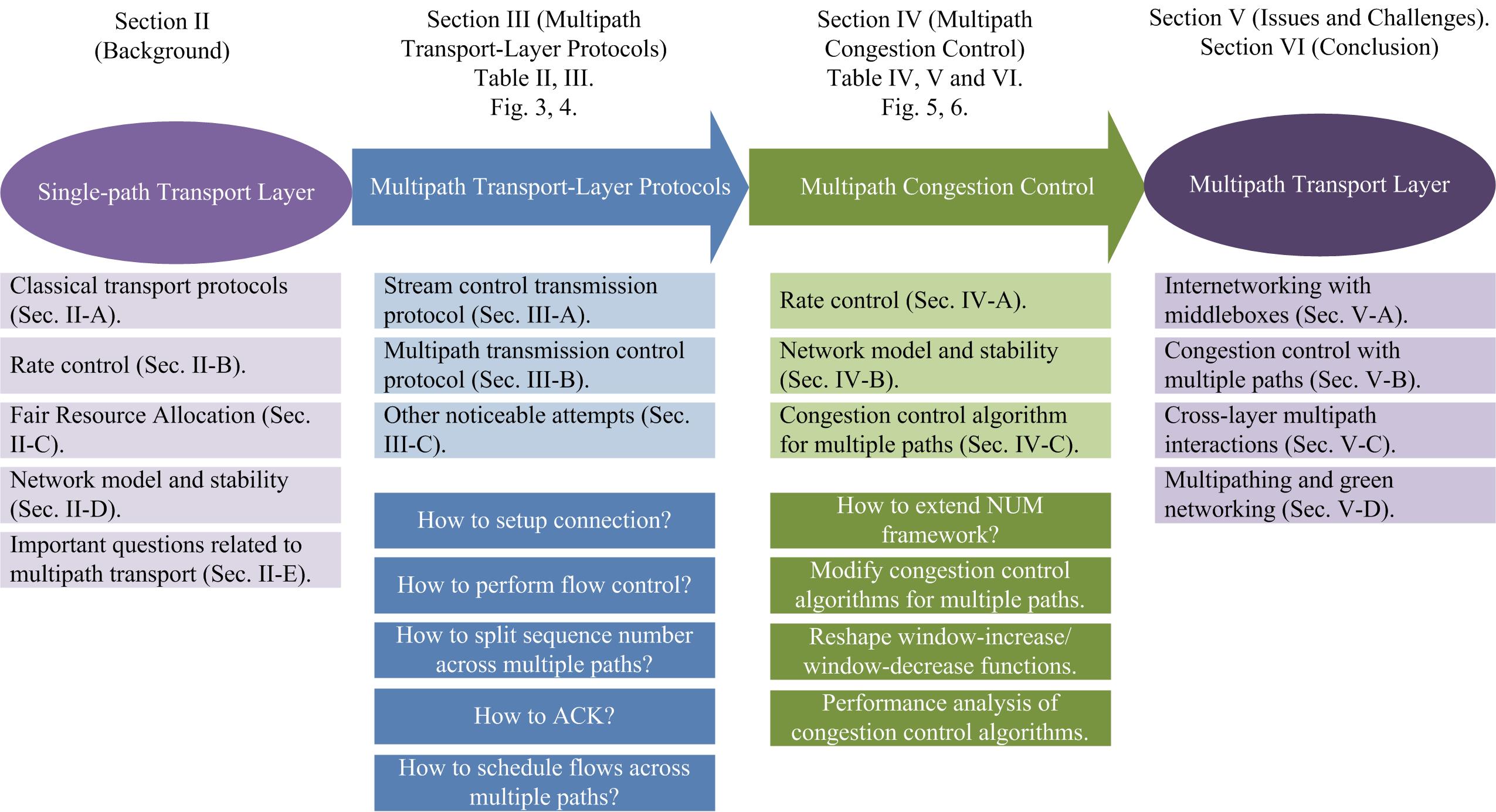}
\caption{Organization of the paper}
\label{taxonomy}
\end{figure*}

While many graceful solutions have been proposed for the aforementioned challenges that will be discussed in our paper, many open research problems still exist in these areas and \textcolor{black}{demand} further research. \textcolor{black}{We will discuss these open research issues in Section \ref{Issues and Challenges}. }

\subsection{Contributions of this Paper}
There \textcolor{black}{exist} a few survey papers that discuss the state-of-the-art in multipathing from the \textcolor{black}{transport-layer} perspective. However, most of them perform a generalized discussion on multipath \textcolor{black}{transport-layer} protocols and do not go into the specifics of protocol design. Table \ref{tab:Related Surveys} analyzes related survey papers \textcolor{black}{in terms of the topics covered in them.} 


To the best of our knowledge, this is the first systematic effort towards explaining the paradigm-shift from single-path to multipath support at the transport layer. The objective of this paper is to benefit the networking community by providing a sound background in transport-layer multipathing. Our contributions are threefold. 
\begin{enumerate}
\item \textcolor{black}{We present the transition from single-path to multipath transport layer by discussing the \textcolor{black}{connection setup}, flow control, sequence number splitting, acknowledgments\textcolor{black}{,} and flow scheduling mechanism of various multipath solutions at the transport layer.}
\item \textcolor{black}{The hallmark of the  transport layer---i.e., congestion control--- has been thoroughly examined in our paper. In particular,}
\begin{itemize}
\item \textcolor{black}{We analyze the mathematical model for rate control with multiple paths in terms of its stability and equilibrium properties.} 
\item \textcolor{black}{Following that, various multipath congestion control algorithms are explained \textcolor{black}{alongside their window-increase/window-decrease} functions  \textcolor{black}{and advantages and disadvantages}.} 
\item \textcolor{black}{Afterwards, these multipath congestion control algorithms are explored for their performance in terms of TCP-friendliness and responsiveness.}
\end{itemize}
\item \textcolor{black}{We identify open research issues and challenges in the field of transport-layer multipathing to provide a direction for future research.}
\end{enumerate} 



\subsection{Organization of this paper}
As shown in Fig. \ref{taxonomy}, our paper is organized around explaining the successful journey from single-path to multipath support at the transport layer. In particular, section \ref{Background} gives a background on single-path transport protocols to familiarize the reader with the fundamentals of protocol design and sets the stage for the rest of the paper by posing the important design challenges that arise with the use of multiple paths. \textcolor{black}{Section \ref{Multipath Transport Layer Protocols} presents the current state-of-the-art for multipath transport protocols. Note that light blue color in Fig. \ref{taxonomy} indicates the sequence of discussion in Section \ref{Multipath Transport Layer Protocols} and dark blue highlights the design issues that are covered in this section.} \textcolor{black}{Section IV performs an in-depth study on multipath congestion control, the course of discussion and important design concerns are shown by light and dark green colors respectively.}
Section \ref{Issues and Challenges} presents the open research challenges and issues related to multipath transport protocols and finally the paper concludes in section \ref{Conclusion} with a summarizing discussion. 

\section{Background}
\label{Background}
\begin{quote}
\textcolor{black}{\textit{``The journey of a thousand miles begins with a single step."--- Lao Tzu}}
\end{quote}
This section aims to provide the reader with a background on single-path transport-layer protocols. \textcolor{black}{This is followed by a discussion on the relationship between rate control \footnote{Note that rate control refers to controlling transmission rate at both source hosts and routers. Congestion control refers to controlling data rate at source hosts only while flow control prevents the sender from over running the capacity of receivers.} and fair resource allocation. Fluid flow models, which are used for fair resource allocation, and the stability of the networks that deploy these methods is also investigated}. At the end of this section we present some of the design questions that are important to consider while designing a multipath transport-layer protocol.

\subsection{Classical Transport Protocols}
\label{Classical Transport Protocols}
User datagram protocol (UDP) \textcolor{black}{\cite{postel1980rfc}} and reliable transmission control protocol (TCP) \textcolor{black}{\cite{postel1981transmission}} are the two classical transport layer protocols. These protocols are application specific; UDP, due to the unreliability feature is suitable for delay sensitive applications while TCP is used for loss sensitive applications as it provides  reliable data delivery. However, the \textcolor{black}{transport-layer} protocols need some additional functionality from application layer to support real-time applications. This has led to the development of real-time transport (RTP) \textcolor{black}{\cite{jacobson2003rtp}} protocol. Next, we provide a comprehensive overview of these three protocols.

UDP provides end-to-end best effort data delivery service by using a single-path between the source and destination. 
UDP packets are bound to the single IP address used by the end-points because UDP packets are decoded on a 5-tuple comprising the IP addresses of the end-points, port identifiers\textcolor{black}{,} and the protocol being used. UDP does not use sequence numbers and packets are expected to arrive as a continuous stream. In case of congestion or link failure, the packets are lost. UDP is a simple addition to the best-effort service model of IP and does not has any flow control and congestion control mechanisms \cite{peterson2007computer}. 

TCP is a reliable end-to-end transport protocol that uses a single-path for data transmission. TCP packets, like UDP, are decoded on a 5-tuple basis, thereby bounding TCP connections to the single IP addresses of the end-points. \textcolor{black}{Connection is established in TCP using a 3-way handshake---i.e., SYN, SYN+ACK,  and ACK.} TCP uses sequence numbers for detecting lost packets and reconstruction of data at the receiver. \textcolor{black}{TCP guarantees reliability by having TCP receiver send acknowledgments (ACKs) to the sender after a successful transmission}. Network congestion and transmission errors can result in a packet/ ACK loss. In such a case, TCP sender waits for three duplicate ACKs, after which retransmission occurs. \textcolor{black}{Retransmissions are essential for recovering lost data, but they also utilize precious bandwidth that otherwise could have been used for transmitting new data. Thus, it is important to avoid unnecessary packet retransmission as it has the potential of wasting useful bandwidth}. TCP selective ACK (SACK) \cite{floyd2000extension}, delayed ACK \cite{braden1989requirements}\textcolor{black}{,}  and cumulative ACK have been proposed in literature to minimize the problem of unnecessary retransmissions.    

Flow control and congestion control are two  different mechanisms with different objectives. Flow control prevents the sender from over running the capacity of receivers while congestion control keeps the hosts from injecting too much traffic in the network. \textcolor{black}{Flow control uses the concept of a sliding receive window to define the amount of data that a sender is allowed to send to the receiver without receiving an ACK.} \textcolor{black}{For congestion control, source calculates the size of congestion window by making use of congestion control algorithms, which exploit information about packet loss and packet delay that is readily  available at source hosts}. \textcolor{black}{TCP implements flow control and congestion control} by having each source set its window size as the minimum of congestion window (as maintained at the sending side) and receive window (advertised by the receiver of a data communication).

TCP was traditionally regarded as not suitable for live streaming as its retransmission and backoff might lead to long delays. Although, recent studies \cite{sripanidkulchai2004analysis, van2002streaming, wang2001empirical} have shown that a significant portion of the Internet traffic uses HTTP/ TCP, but this fact was not established at the time of development of RTP, which was developed for providing end-to-end real-time data delivery service using single-path. Some functions provided by RTP are part of the transport layer while others belong to application layer. 

\textcolor{black}{\textcolor{black}{RTP is used to transport real-time data} and consists of an associated RTP control protocol (RTCP) to observe QoS by taking feedback from receivers. RTP typically runs on top of UDP/ IP but any other transport protocol can also be used.  Congestion control in RTP can vary depending on the application demands. A typical congestion mechanism is to adapt data rate based on RTCP feedback.}

After discussing the basics of classical transport protocols, \textcolor{black}{we now move towards the main feature of the  transport layer, i.e., rate control}. While UDP does not implement rate control, RTP's rate control is dependent on specific applications.  TCP, on the other hand, comprises the major portion of the Internet traffic and implements rate control. That is why we next discuss rate control in Internet from TCP's perspective.  




\subsection{Rate Control}
\label{rate control}
The Internet is a vast collection of communication links and resources that are shared by diverse sources. The Internet is managed in a distributed fashion without any single governing entity. Congestion occurs when the arrival rate at a resource outstrips its service capacity. This leads to increased queuing and buffering delays \textcolor{black}{that} eventually results in buffer overflows and dropped packets. 

The need of congestion control was first felt in the 1980s when the Internet had its first \emph{congestion collapse}. The initial version of the TCP protocol then used on the Internet did not have any mechanisms for rate control to keep in account the congestion of the network. Although TCP did implement flow control that determined the size of the congestion window with respect to the receiver's processing capacity \textcolor{black}{but---without} appropriate throttling of sending rates in the case of network congestion, networks throughput could dwindle down to crawling speeds. This had led to the widely publicized meltdown of Internet performance termed as a `congestion collapse'. Van Jacobson proposed
the congestion control algorithms for TCP that  allowed TCP to implement congestion control \cite{jacobson1988congestion}.

Since then various congestion control algorithms have been proposed in literature for TCP including proposals that rely on (i) packet loss (such as TCP Reno \cite{padhye2000modeling}), (ii) delay (such as TCP Vegas \cite{brakmo1995tcp}), (iii) combination of loss and delay (such as TCP Illinois \cite{liu2008tcp}). Almost all the  congestion control algorithms are developed on the principle of increasing the size of the congestion window when an ACK arrives and decreasing the size if a loss occurs. This way the size of the window is dynamically adjusted depending on level of congestion in the network.

Congestion can also be controlled at the interim routers by allowing routers to mark (or even drop) packets based on the queue length thereby indicating congestion to the sender. For this purpose, routers use active queue management (AQM) technology to actively manage their queues. AQM can be implemented using various algorithms including drop tail \cite{gevros2001congestion}, random early detection (RED) \cite{floyd1993random}, random early marking (REM) \cite{athuraliya2000random}, active virtual queue (AVQ) \cite{kunniyur2001analysis}, controlled  delay (CoDel) \cite{nichols2012controlling}.

Since the Internet has a decentralized architecture (with each user being in control of its transmission rate), an important question is: how do we devise appropriate incentives for users to remain fair to each other? More importantly, does fairness simply means equal? We address the subject of these questions in the next subsection.

\subsection{Fair Resource Allocation}
\label{Fair Resource Allocation}
In order to understand the problem of fair resource allocation, consider an example scenario where two users A and B are sharing a resource that has a capacity of 12 Mbps. Suppose user A requires 8 Mbps and user B requires 4 Mbps. Fairness (conceived in terms of the ideal of equal allocation) implies that the capacity be equally shared between the two users---i.e., each user should get 6 Mbps. With this kind of resource allocation, user A gets less satisfaction as it gets less than what it requires while capacity is wasted with user B  (who only needs 4 Mbps). Since the link capacity has the potential to fully satisfy the needs of both users thus a better way is to allocate resources according to the needs of the two users. This implies allocating 8 Mbps to user A and 4 Mbps to user B. This unequal resource allocation is made fair by introducing the concept of \emph{shadow price} \cite{kelly1998rate}. Shadow price is a price that each user has to pay for the amount of resources it utilizes. Price associated with a resource in the network is basically a congestion measure and has different interpretations in different protocols (loss probability in TCP Reno and queuing delay in TCP Vegas). 

\textcolor{black}{The problem of resource allocation and sharing between multiple stakeholders is studied in the technique of game theory.  Nash \cite{nash1950bargaining} studied the problem of resource allocation in a non cooperative game-theoretic setting based on which Nash was awarded the Nobel Prize. In a non-cooperative game, different interacting entities or players are said to be in a `Nash equilibrium' state if no one player can gain if it changes its strategy when all the other players keep their strategies constant. It is important to note here that it is not always the case that Nash Equilibrium provides a fair or globally optimal solution. This is evident by the Nash equilibrium state achieved in the famous `prisoner’s dilemma' game.}


Various attempts have been made in literature to axiomatically characterize what might constitute a ``fair''  allocation of resources in the context of networking \cite{chiang2010fairness}. \emph{Proportional fairness} \cite{kelly1997charging} implies that any change in the distribution of rates, results in the sum of proportional changes being negative. In \emph{max-min fairness} \cite{thomas1980decentralized}, all users get same share at bottleneck link. \emph{TCP fairness} \footnote{In this paper, we interchangeably use the terms TCP fairness and TCP-friendliness as both imply the same thing.} measures fairness of other Internet traffic (e.g., UDP) to TCP flows. Since best effort traffic is unrespovsive to packet drop rate (due to the absence of congestion control algorithms), it runs the risk of being unfair to TCP flows \cite{floyd1999promoting}. 

The concept of \emph{$\alpha$-fairness} proposed by Mo and Walrand \cite{mo2000fair} incorporates a single parameter family of objectives (parameterized by $\alpha$) that generalizes the notions of proportional fairness, max-min fairness\textcolor{black}{,} and TCP fairness. In particular, $\alpha$-fairness subsumes both max-min fairness ($\alpha$ = $\infty$), TCP fairness ($\alpha$ = $2$), and proportional fairness ($\alpha$ = 1) as special cases. \textcolor{black}{The above-mentioned fairness concepts are also applicable to multipath traffic. However, our paper concentrates on the problem of TCP fairness with multipath traffic, an issue that will be considered in detail in section \ref{Congestion Control Algorithms for Multiple Paths}}. 

Kelly \cite{kelly1997charging} realized the close relationship between fair resource allocation and congestion control. Based on that, \textcolor{black}{Kelly} formulated a mathematical model for charging a source host for a particular transmission rate in order to ensure fairness. This is the topic of discussion in the next subsection.

\subsection{Network Model and Stability}
\label{Network Model and Stability}
Researchers have realized that congestion control algorithms are essentially distributed resource allocation algorithms that are implicitly solving a global optimization problem and performing NUM \cite{kelly1998rate, kelly2000models, kelly2001mathematical,  kelly1997charging, shakkottai2008network}.

The utility maximization problem in a network is to maximize the aggregate utility of source nodes subject to link constraints.
\begin{equation}
\label{Equation1}
\max \sum_{s \in S} U_s(x_s) \quad \text{subject to} \quad y_l \leq c_l
\end{equation}

The above equation is motivated as follows: each user $s$ transmits at rate $x_s$ on a route composed of $l$ number of links. The satisfaction that the user obtains from its transmission rate is expressed by the user's utility function $U_s(x_s)$. 
The aggregate transmission rate of different sources $y_l$ accessing the link is constrained by link capacity $c_l$. 

Utility maximization in Internet is a distributed problem as each source and link has to act as a controller and solve the optimization problem using only local information. For facilitating distributed computing, the optimization problem (\ref{Equation1}) is formulated into an appropriate form that can be primal, dual\textcolor{black}{,} and primal-dual \cite{kelly2005stability, kelly1998rate, kelly2009resource,  srikant2004mathematics}. Here, we present only the primal-dual formulation of the optimization problem (\ref{Equation1}).
\begin{equation}
\label{Equation12}
\dot{x_r}= k_r(x_r)(U'_r(x_r)-q_r)
\end{equation}
\begin{equation}
\label{Equation13}
 \dot p_l = h_l(y_l-c_l)^+_{p_l} 
\end{equation}

where
\begin{itemize}
\item where $p_l$ is the price associated with link $l$.
\item $q_r$ is the aggregate price of a route $r$, such that $q_r = \sum_{l \in r} p_l$.
\item $k_r$ and $h_l$ are positive scalar quantities.
\item $(a)^+_x$ means $a$ for $x > 0$ and $max (a,0)$ for $x = 0$.
\end{itemize}

As mentioned before, congestion can be either controlled at the source hosts using TCP congestion control algorithms or at \textcolor{black}{the} interim routers using AQM technology. Primal-dual formulation uses congestion control at \textcolor{black}{the} source hosts and \textcolor{black}{at the} interim routers. In particular, the sources in the network adjust their rates $x_r$ using TCP congestion control algorithms depending on the feedback from the network which is in terms of aggregate link prices $q_r$. A congested link has a higher price than an uncongested link. The links, on the other hand, use AQM technology to adjust their prices $p_l$ based on aggregate source rates $y_l$. 

The convergence and uniqueness of fluid flow models to an equilibrium solution is essential and is ensured by associating a strictly concave utility function with sources. The dynamic properties of the fluid flow models (presented by equations (\ref{Equation12}) and (\ref{Equation13})) are studied using the machinery of control theory to understand the stability and optimality of rate control algorithms. Kelly et al. established the stability of fluid flow models by showing that an appropriate formulation (primal, dual form) of optimization problem expressed in equation (\ref{Equation1}) provides a Lyapunov function for the system defined by the rate control algorithm \cite{kelly1998rate}.

Flow arrivals/ departures, propagation delays\textcolor{black}{,} and queuing delays all cause the network to enter in a transient state. Fluid flow models help us to study whether a network in transient state will converge to an equilibrium state. Equilibrium state is the point at which the network has a desired behavior in terms of throughput and delay. 

\textit{Limitations of Fluid Models:} The fluid flow models do not take into account the randomness \textcolor{black}{in the network due to random} packet arrivals. \textcolor{black}{The} actual behavior of the network is always stochastic and \textcolor{black}{Markov models are} used for stochastic modeling. \textcolor{black}{Markov models specify network state in the next RTT based on information about network parameters (e.g., the congestion window) in the current RTT.} The use of stochastic models has been emphasized by arguing that fluid-based models are restrictive in terms of the buffer capacity and the link utilization \cite{eun2007limitation}.

However, stochastic models are very difficult to analyze. \textcolor{black}{Since rate control is largely dominated by congestion feedback signals and as fluid flow and stochastic models achieve same equilibrium point (in case of multiple flows) so, fluid models are more frequently used for modeling network behavior \cite{low2002internet}.} Translation of \textcolor{black}{the} fluid flow models in practice requires care but is perfectly feasible \cite{key2011path}.
\\

Until this point, classical transport control protocols and use of rate control for fair resource allocation have been presented. The network model for rate control was also discussed \textcolor{black}{along with} an analysis \textcolor{black}{of} its stability. Now, a natural followup question is: how the principles of single-path transport layer can be extended to multiple paths? What design challenges arise in such a scenario? The next subsection deals with these questions. 

\subsection{Important Questions Related to Multipath Transport}
\label{Important Question Related to Multipath Transport}
As mentioned before, classical transport control protocols do not have multihoming support. This implies that if the transmission path between the source-destination pair becomes unavailable, then communication will be blocked forever or until the availability of that particular path. In general, it is always better to diversify and not put all eggs in one  basket. Researchers in agreement with this  philosophy, started efforts towards extending the principles of single-path transport layer to multiple paths. 

This extension raised a number of design questions and challenges, all of which were carefully addressed by the protocol designers. Before getting into the details of these design issues, let us first define the notion of a  flow, a subflow, a connection\textcolor{black}{,}  and a path. \textcolor{black}{A flow is simply the data to be transmitted. It is associated with a connection that can have multiple paths.} This flow, belonging to a particular connection can  split into subflows,  which are then transmitted across multiple paths. On arriving at the destination, these subflows combine to reconstruct the original message. Now, we are in a position to formally start our discussion on multipath transport layer by broadly presenting \textcolor{black}{nine} design issues one after another.  \\ 

\begin{figure} 
\centering
\includegraphics[width=8cm]{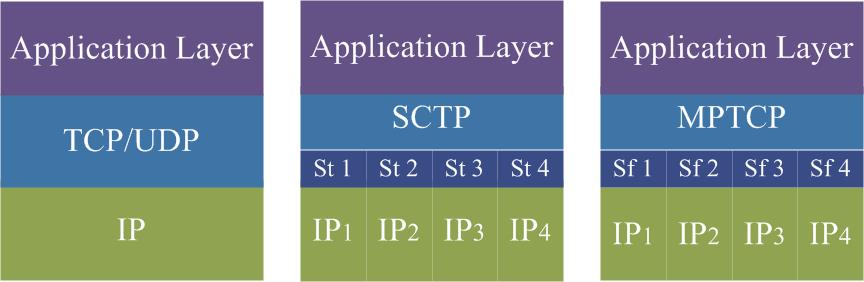}
\caption{Architecture of UDP, TCP, SCTP and MPTCP }
\label{fig9}
\end{figure}

\begin{enumerate}
\item {\textcolor{black}{\textit{Connection Establishment}}: How to establish connection with multiple paths? TCP uses a 3-way handshake for establishing connection between endpoints. So, can we tweak this 3-way handshake for setting up connection between a source-destination pair that have multiple paths or do we need a new scheme?} \\

\item{\textcolor{black}{{\textit{Flow Control}}: Flow control can be performed on a per connection or per path basis\textcolor{black}{/ per subflow basis (per path and per subflow are interchangeably used throughout the paper)?} Connection level flow control maintains a shared buffer for all paths while per path flow control keeps a separate buffer for all paths. Now, which one of the two schemes is better and why?}} \\ %

\item {\textcolor{black}{\textit{Sequence Number Splitting}: Classical TCP uses a single sequence space for detecting lost packets and reconstructing  message at the receiver. Is this single sequence space sufficient for multipath transport-layer protocols? Do we need double sequence space, one per path and second per connection? Which one of these ideas is more feasible and practical?}} \\ 

\item \textcolor{black}{\textit{ACK}: 
With multiple paths, are ACKs required at connection level or subflow level or both? When multiple paths with heterogeneous path characteristics are used, then ACKs may arrive at the receiver out of order. So, which mechanism is more suitable for \textcolor{black}{catering out of order delivery problem, SACKs, delayed ACKs or cumulative ACK?}} \\

\item \textcolor{black}{\textit{Flow Scheduling}: Flow scheduling is a mechanism that is not required with single-path transport layer. However, with multiple paths there is a need to schedule flow across multiple paths. The most basic flow scheduling technique in networking is round robin. So, is this scheme good enough or do we need more sophisticated flow scheduling algorithms? The heterogeneous path  characteristics play a key role in designing flow scheduling algorithms for multipath transport layer.} \\

\item \textcolor{black}{\textit{NUM Framework}:  \textcolor{black}{Can the} utility maximization framework for single-path transport layer be extended to multiple paths? \textcolor{black}{Is the resulting model stable and doest the model converge to an equilibrium state?}} \\ 

\item \textcolor{black}{\textit{Congestion Control}}: 
Just like flow control, congestion control can also be performed on a per path or per connection basis. \textcolor{black}{Specifically, the terms \emph{uncoupled}  (per path basis) and \emph{coupled} \footnote{Note that the term coupled is used to refer to all the coupled congestion control algorithms (such as fully coupled, semi-coupled, linked increase adaptation, etc) while  \textit{fully coupled} refers to the fully coupled congestion control algorithm that is used by MPTCP. More details on this topic is provided in Section \ref{Congestion Control Algorithms for Multiple Paths}.} (per connection basis) are used with congestion control. \textcolor{black}{This intrigues the question: which one of them is better and easy to implement?}} \\ 

\item \textcolor{black}{\textit{Window-Increase/Window-Decrease function}: 
Are the \textcolor{black}{window-increase/window-decrease} functions of regular TCP applicable to multipath scenario? If not, which other algorithms have been proposed and what are their \textcolor{black}{window-increase/window-decrease} functions?} \\

\item \textcolor{black}{\textit{TCP-Friendliness \& Responsiveness}: As already mentioned in section \ref{Challenges in Implementing Multipath Transport Layer}, fairness is an issue of paramount importance. Exploring the performance of congestion control algorithms in terms of TCP-friendliness and responsiveness to changes in network conditions (e.g., congestion level) is essential before their deployment. Based on this performance analysis, a decision regarding which particular rate control algorithm to use in practice is made.} \\
\end{enumerate}

\textcolor{black}{A quick glance at the above-mentioned nine points reveals that the first five design questions are related to protocol specifications in general while the last four are explicitly related to congestion control with multiple paths. This is the logical basis for discussing first five questions in Section \ref{Multipath Transport Layer Protocols} and last four in Section \ref{Multipath Congestion Control}}.

\begin{table*}[!ht]
\centering
\scriptsize
\caption{Multipath Transport Layer Protocols}
\label{tab:SCTP and its variants}
\begin{tabular}{p{2cm}p{0.5cm}p{1.8cm}p{1.8cm}p{2.4cm}p{1cm}p{2cm}p{1.5cm}p{1.6cm}}
\hline

\cellcolor[HTML]{EFEFEF}\textbf{\emph{Transport Protocol}} & \cellcolor[HTML]{EFEFEF}\textbf{\emph{Year}} & 
\cellcolor[HTML]{EFEFEF}\textbf{\emph{Standardization}} &
\cellcolor[HTML]{EFEFEF}\textbf{\emph{\textcolor{black}{Connection Setup}}}  &
\cellcolor[HTML]{EFEFEF}\textbf{\emph{Flow Control}}  &
\cellcolor[HTML]{EFEFEF}\textbf{\emph{Sequence Space}}  &
\cellcolor[HTML]{EFEFEF}\textbf{\emph{ACK Mechanism}}&
\cellcolor[HTML]{EFEFEF}\textbf{\emph{Flow Scheduling}} 
&
\cellcolor[HTML]{EFEFEF}\textbf{\emph{Congestion Control}}   
\\ 
\hline

UDP \cite{postel1980rfc} & 1980 & IETF RFC 768 & \textcolor{black}{N/A} & N/A & N/A & N/A & N/A & \textcolor{black}{N/A}\\
TCP \cite{postel1981transmission} & 1981 & IETF RFC 793 & \textcolor{black}{3-way handshake} & Per connection & Single & \textcolor{black}{Cumulative ACK}, SACK, Delayed ACK & N/A & \textcolor{black}{Uncoupled} 
\\
\multicolumn{8}{l}{\textbf{\underline{\emph{SCTP and its variants}}}}\\

SCTP \cite{stewart2000stream, stewart2007stream} & 2000$,$ 2007 & IETF RFC \textcolor{black}{4960} & \textcolor{black}{4-way handshake} & Per association & Single & SACK & N/A & \textcolor{black}{Uncoupled}  
\\

PR-SCTP \cite{stewart2002sctp, stewart2004rfc} & 2002$,$ 2004 & IETF RFC 3758 & \textcolor{black}{4-way handshake} & Per association & Single & SACK & Not specified & \textcolor{black}{Uncoupled}  
\\

BA-SCTP \cite{argyriou2003bandwidth} & 2003 & Not standardized & \textcolor{black}{4-way handshake} & Per association & Single &  SACK & WRR & \textcolor{black}{Uncoupled with SBD} 
\\

DAR-SCTP \cite{stewart2003stream, stewart2005stream, maruyama2007stream} & 2003$,$ 2005$,$ 2007 & IETF RFC 5061 & \textcolor{black}{4-way handshake} & Per association & Single & SACK & Not specified & \textcolor{black}{Uncoupled}  
\\

W-SCTP \cite{casetti2004westwood} & 2004 & Not standardized & \textcolor{black}{4-way handshake} & Per path (sender side), Per association (receiver side) & Single & SACK & EDPF & \textcolor{black}{Uncoupled}
\\ 

LS-SCTP \cite{abd2004ls, abd2004improving} & 2004$,$ 2010$,$ 2013 & Internet Draft \cite{becke2010load, amer2013load} & \textcolor{black}{4-way handshake} & Per association & Double & SACK & WRR & \textcolor{black}{Uncoupled}
\\

m-SCTP \cite{koh2005mobile, riegel2007mobile} & 2005$,$ 2007 & Internet Draft \cite{riegel2007mobile} & \textcolor{black}{4-way handshake} & Per association & Single & SACK & Not specified & \textcolor{black}{Uncoupled}  
\\

CMT-SCTP \cite{iyengar2006concurrent} & 2006 & Not standardized & \textcolor{black}{4-way handshake} & Per path & Single & SACK \& Delayed ACK & WRR & \textcolor{black}{Uncoupled with SBD} 
\\

WiMP-SCTP \cite{huang2007wimp} & 2007 & Not standardized & \textcolor{black}{4-way handshake} & Per association & Single & SACK & WRR & \textcolor{black}{Uncoupled} 
\\ 

cmpSCTP \cite{liao2008cmpsctp} & 2008 & Not standardized & \textcolor{black}{4-way handshake} & Per association & Double & SACK & WRR & \textcolor{black}{Uncoupled} 
\\

mSCTP-CMT \cite{budzisz2009concurrent} & 2009 & Not standardized & \textcolor{black}{4-way handshake} & Per association & Single & SACK & WRR & \textcolor{black}{Uncoupled with SBD} 
\\

\\

FPS-SCTP \cite{mirani2010data} & 2010 & Not standardized & \textcolor{black}{4-way handshake} & Per association & Single & SACK & FPS & \textcolor{black}{Coupled} 
\\

WM$^2$-SCTP \cite{yuan2010extension} & 2010 & Not standardized & \textcolor{black}{4-way handshake} & Per path & Triple & SACK & \textcolor{black}{QoS \& WRR} & \textcolor{black}{Uncoupled with SBD} 
\\

\multicolumn{8}{l}{\textbf{\underline{\emph{MPTCP and its variants}}}}\\

MPTCP \cite{ford2011architectural, raiciu2012hard, ford2013tcp}  & \textcolor{black}{2011-2013} & IETF RFC 6182 & \textcolor{black}{3-way handshake} & Per connection & Double & SACK \& DSN-ACK & WRR & \textcolor{black}{Coupled} 
\\

NC-MPTCP \cite{li2012network} & 2012 & Not standardized & \textcolor{black}{3-way handshake} & Per connection & Double & SACK \& DSN-ACK & FPS & \textcolor{black}{Coupled}  
\\

QoS-MPTCP \cite{diop2012qos} & 2012 & Not standardized & \textcolor{black}{3-way handshake} & Per connection & Double & SACK \& DSN-ACK & \textcolor{black}{QoS \& WRR} & \textcolor{black}{Coupled} 
\\ 

MPTCP \cite{li2013delayed} & 2013 & Not standardized & \textcolor{black}{3-way handshake} & Per connection & Double & New Delayed ACK & WRR & \textcolor{black}{Coupled} 
\\

MPTCP/OpenFlow \cite{van2013experiences} & 2013 & Not standardized & \textcolor{black}{3-way handshake} & Per connection & Double & SACK \& DSN-ACK & WRR & \textcolor{black}{Coupled} 
\\

A-MPTCP \cite{coudron2013cross} & 2013 & Not standardized & \textcolor{black}{3-way handshake} & Per connection & Double & SACK \& DSN-ACK & WRR & \textcolor{black}{Coupled} 
\\

CWA-MPTCP \cite{zhou2013goodput} & 2013 & Not standardized & \textcolor{black}{3-way handshake} & Per connection & Double & SACK \& DSN-ACK & FPS & \textcolor{black}{Coupled} 
\\

SC-MPTCP \cite{li2013tolerating} & 2014 & Not standardized & \textcolor{black}{3-way handshake} & Per connection & Double & SACK \& DSN-ACK & FPS & \textcolor{black}{Coupled}  
\\

Yang and Amer \cite{yang2014using} & 2014 & Not standardized & \textcolor{black}{3-way handshake} & Per connection & Double & SACK \& DSN-ACK & FPS & \textcolor{black}{Coupled}  
\\

FMTCP \cite{cui2015fmtcp} & 2015 & Not standardized & \textcolor{black}{3-way handshake} & Per connection & Double & SACK \& DSN-ACK & FPS & \textcolor{black}{Coupled} 
\\

Le and Bui \cite{le2015forward} & 2015 & Not standardized & \textcolor{black}{3-way handshake} & Per connection & Double & SACK \& DSN-ACK & FDPS & \textcolor{black}{Coupled}  
\\
\\
\multicolumn{7}{l}{\textbf{\underline{\emph{Other Noticeable Attempts}}}}\\

R-MTP \cite{magalhaes2001transport} & 2001 & Not standardized & \textcolor{black}{3-way handshake} & Per connection  \& maximum rate & Single &  SACK & WRR & \textcolor{black}{Uncoupled} 
\\

Lee et al. \cite{lee2002improving} & 2002 & Not standardized & \textcolor{black}{3-way handshake} & Per connection & Single & Delayed ACKs & Not specified & \textcolor{black}{Uncoupled}  
\\

pTCP \cite{hsieh2002ptcp, hsieh2005transport} & 2002$,$ 2005 & Not standardized & \textcolor{black}{3-way handshake} & Per connection & Double & SACK & WRR & \textcolor{black}{Uncoupled} 
\\

R$^2$CP \cite{hsieh2003receiver, kim2005receiver} & 2003$,$ 2005 & Not standardized & \textcolor{black}{3-way handshake} & Per connection & Double & SACK & EDPF & \textcolor{black}{Uncoupled}  
\\

Cetinkaya and
Knightly \cite{cetinkaya2004opportunistic} & 2004 & Not standardized & \textcolor{black}{3-way handshake} & Per connection & Single & SACK & OMS & \textcolor{black}{Uncoupled} 
\\

mTCP \cite{zhang2004transport} & 2004 & Not standardized & \textcolor{black}{3-way handshake} & Per connection & Single & SACK & WRR & \textcolor{black}{Uncoupled with SBD}  
\\

M-TCP \cite{chen2004multipath} & 2004 & Not standardized & \textcolor{black}{3-way handshake} & Per connection & Single & \textcolor{black}{Cumulative ACK} & Not specified & \textcolor{black}{Uncoupled} 
\\

M/TCP \cite{rojviboonchai2004evaluation} & 2004 & Not Standardized & \textcolor{black}{3-way handshake} & Per connection & Single & Duplicated \& Delayed ACK & WRR & \textcolor{black}{Uncoupled with SBD}  
\\ 

R-M/TCP \cite{rojviboonchai2005rm} & 2005 & Not standardized & \textcolor{black}{3-way handshake} & Per connection & Single & Acked Info List for each path & WRR & \textcolor{black}{Uncoupled with SBD} 
\\

cmpTCP \cite{sarkar2006concurrent} & 2006 & Not standardized & \textcolor{black}{4-way handshake} & Per connection & Single & SACK & WRR & \textcolor{black}{Uncoupled}  
\\

cmpTCPW \cite{sarkar2006qrp04} & 2006 & Not standardized & \textcolor{black}{4-way handshake} & Per connection & Single  & SACK & WRR & \textcolor{black}{Uncoupled}
\\

cTCP \cite{dong2007concurrency} & 2007 & Not standardized & \textcolor{black}{3-way handshake} & Per connection & Single & Duplicated ACK classifier & WRR & \textcolor{black}{Coupled} 
\\

MPLOT \cite{sharma2008mplot, sharma2012transport} & 2008$,$ 2012 & Not standardized & \textcolor{black}{3-way handshake} & Per connection & Single & SACK and cumulative ACK & EDPF & \textcolor{black}{Uncoupled}  
\\

\hline
\end{tabular}
\end{table*}

\section{Multipath Transport Layer Protocols}
\label{Multipath Transport Layer Protocols}
\begin{quote}
\textcolor{black}{
\textit{``Variety's the very spice of life, that gives it all its flavor."--- William Cowper}}
\end{quote}

\textcolor{black}{In harmony with Cowper's thoughts, researchers have recommended a huge variety of multipath transport layer protocols, which are briefly overviewed in this section.} 
Stream control transmission protocol (SCTP) and MPTCP are the two most prominent efforts towards transport-layer multipathing. In order to establish the general idea for multipath transport layer, we first refer the reader to Fig. \ref{fig9}. \textcolor{black}{As evident from the figure, both SCTP and MPTCP have multihoming support (with multihoming, the end host is equipped with multiple network interfaces and IP addresses).} SCTP uses the concept of streams (\emph{St} in Fig. \ref{fig9} indicates streams) while MPTCP uses the concept of subflows (\emph{Sf} in Fig. \ref{fig9} represents subflows), both of them will be discussed in detail in the subsequent subsections. After explaining these two milestones, several other noticeable attempts towards the use of multiple paths at the transport layer will be presented. 

\textcolor{black}{\textcolor{black}{Before starting} an exhaustive discussion on the technicalities of multipath transport-layer protocols, let us first organize the \textcolor{black}{connection setup,} flow control, sequence number splitting, ACK, flow scheduling\textcolor{black}{, and congestion control \footnote{A detailed discussion on congestion control follows in Section \ref{Multipath Congestion Control}, but it has been included in Table \ref{tab:SCTP and its variants} for the purpose of saving space.} mechanism for various multipath \textcolor{black}{transport-layer} protocols in Table \ref{tab:SCTP and its variants} to give a clear picture to the reader.} \textcolor{black}{Table \ref{tab:SCTP and its variants} has been adapted from the works presented in \cite{zhuang2012multipath, ramaboli2012bandwidth, habak2013bandwidth}. Nevertheless, its inclusion in our paper is essential to cover the major portions of the story of transition from single-path to multipath support at the transport layer. Note that our tale is not complete because some other key aspects (such as packet header format, connection tear down, and buffer management) have not been covered in our paper.}}

\subsection{Stream Control Transmission Protocol (SCTP)}
\label{SCTP}
The first noticeable work on transport-layer multipathing is SCTP \cite{stewart2000stream}, which is a reliable transport-layer protocol that combines features of classical TCP (e.g., error detection, retransmission, and window-based congestion control \cite{budzisz2008towards}) with several new ones (such as multihoming and multistreaming \cite{budzisz2008towards,jungmaier2006sctp}). In SCTP, an \emph{association} is formed between multihomed hosts--- \textcolor{black}{however, the concept of SCTP association is much broader than TCP connection. SCTP allows endpoints to exchange a list of transport addresses (i.e., multiple IP addresses together with an SCTP port). Accordingly, all the possible transmission paths are determined by generating combinations of the source-destination addresses using each endpoint's list.}

\begin{figure} 
\centering
\includegraphics[width=9cm,keepaspectratio]{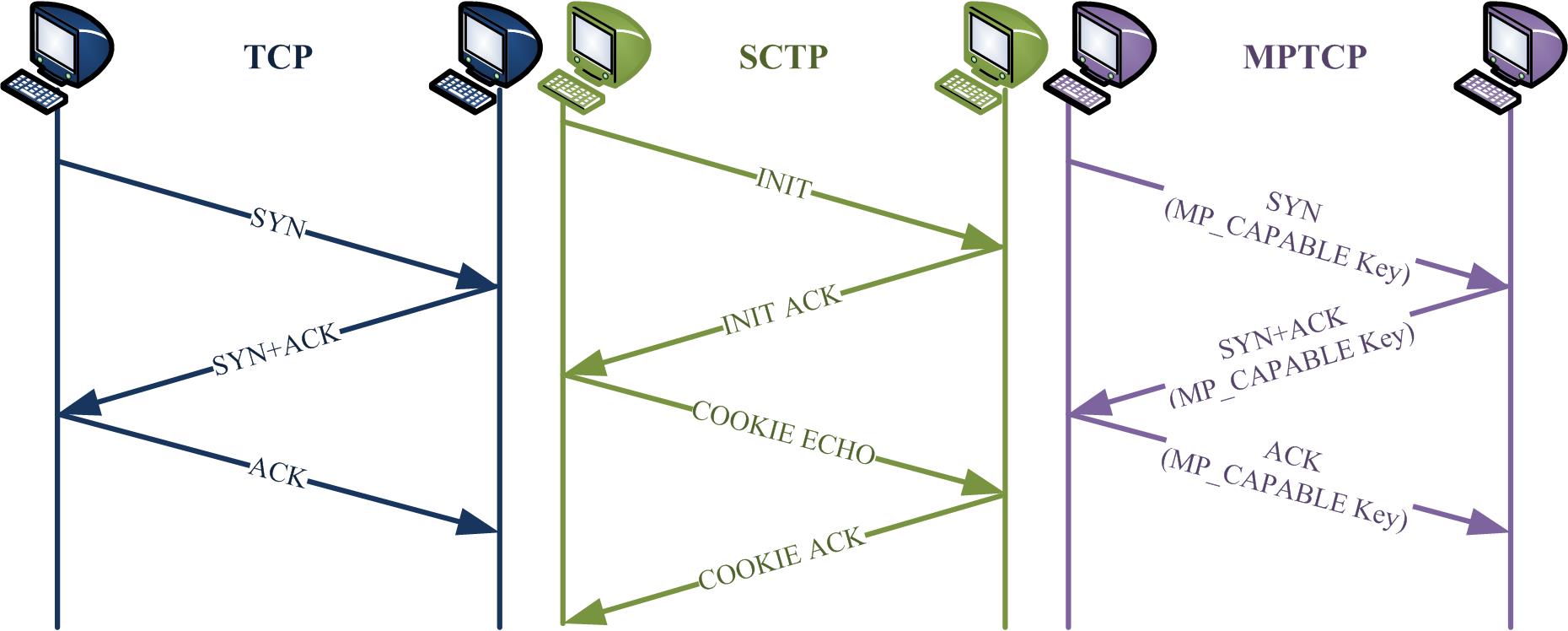}
\caption{Connection Establishment in TCP, SCTP and MPTCP}
\label{Connection Establishment}
\end{figure}

\textcolor{black}{Despite having multiple paths, SCTP uses a single primary path for information transfer while the secondary paths are used for retransmissions or in the case of a primary path failure.} SCTP constantly monitors all the available paths using heartbeat chunks \textcolor{black}{and renders them as active or inactive depending on heartbeat ACK.}

\textcolor{black}{SCTP was designed to use a single-path in order to avoid the problem of fairness with other Internet traffic, in particular with TCP flows. However, load balancing is the key incentive of multipathing. \textcolor{black}{Thus, several efforts towards making SCTP capable of reaping the benefits of bandwidth aggregation were made including bandwidth aggregation based on SCTP (BA-SCTP) \cite{argyriou2003bandwidth}, westwood SCTP (W-SCTP) \cite{casetti2004westwood}, load sharing SCTP (LS-SCTP) \cite{abd2004improving, abd2004ls}, cmpSCTP\cite{liao2008cmpsctp}, concurrent multipath transfer SCTP (CMT-SCTP) \cite{Iyengar2006}, and wireless multipath SCTP (WiMP-SCTP) \cite{huang2007wimp}.}} \textcolor{black}{Next, we present the nuts and bolts of SCTP, i.e., details about the connection establishment, flow control, sequence number splitting, ACK, and flow scheduling mechanisms.}

\textcolor{black}{Connection is setup in SCTP using a 4-way handshake--- INIT, INIT ACK, COOKIE ECHO, and COOKIE ACK. An association setup request is sent via INIT chunk. The \emph{cookie}, which contains information about endpoints (e.g., IP addresses, window size) is transferred between the endpoints through INIT ACK and COOKIE ECHO messages. Finally, COOKIE ACK concludes the connection (reader can look at Fig. \ref{Connection Establishment} for a graphical illustration).}

Flow control in SCTP can \textcolor{black}{either} be performed on a per association or \textcolor{black}{a} per path basis. For association level flow control, a shared buffer is maintained for all paths, while a separate buffer for each path is maintained if flow control has to be performed on a per path basis. \textcolor{black}{SCTP} implements \textcolor{black}{flow control on an association basis, however certain SCTP variants perform flow control on a per path basis \cite{casetti2004westwood, iyengar2006concurrent,yuan2010extension}.}

SCTP, like TCP\textcolor{black}{,} uses a single sequence space called transmission sequence number (TSN) for detecting lost packets and reconstruction of data at the receiver. However, some variants of SCTP use double (i.e., one per path called  path sequence number (PSN) and second per association called association sequence number (ASN)) or triple sequence space (i.e., PSN, ASN, and third per flow called flow sequence number (FSN)) for the same purpose \textcolor{black}{\cite{abd2004improving, abd2004ls,liao2008cmpsctp,yuan2010extension}}.

When multiple paths having different path characteristics are used, then they can cause out-of-order data delivery at the receiver. This could lead to unnecessary fast retransmissions that are not due to lost but delayed packets. SCTP uses SACKs \textcolor{black}{ \cite{stewart2000stream, stewart2007stream}} and delayed ACKs \textcolor{black}{ \cite{iyengar2006concurrent}} to cater for this problem. 

\textcolor{black}{The problem of unnecessary fast retransmissions can also be reduced by designing appropriate flow scheduling mechanisms. \textcolor{black}{Round robin is the simplest flow scheduling technique but it does not take into account the heterogeneous path characteristics (e.g., available bandwidth, RTT, and size of congestion window)}}. \textcolor{black}{Thus weighted round robin (WRR), earliest delivery path first (EDPF) \textcolor{black}{\cite{casetti2004westwood},}  and forward prediction scheduling (FPS) \textcolor{black}{ \cite{mirani2010data}} have been proposed to reduce packet reordering at the receiver. Let us briefly overview the afore-named flow scheduling techniques.} \textcolor{black}{WRR performs flow splitting with the aim of maximizing overall throughput across multiple paths. This scheme is most beneficial when the available paths have similar characteristics. 
EDPF schedules packets based on the estimated delivery time of packets at the receiver. FPS reduces data reordering at the receiver by  scheduling data across multiple paths according to the delay incurred on each path, so that data arrived at the receiver retains its order.} \textcolor{black}{As already mentioned, SCTP uses a single-path and does not require any flow scheduling mechanism.} 

\textcolor{black}{\textcolor{black}{In addition to unnecessary fast retransmission, another vital issue is the head of line (HOL) blocking problem.} HOL blocking is a phenomena in which the processing of high sequence number packets by an application is held back due to delay in the arrival of low sequence number packets. \textcolor{black}{SCTP introduced the concept of streams to mitigate the HOL blocking problem.}  \textcolor{black}{A stream is essentially a subflow that enables} SCTP to decouple reliability from message ordering. \textcolor{black}{This decoupling has been made possible by the use of} stream sequence numbers (SSN) that ensure in order delivery within a stream while unordered delivery can occur across streams. As a result, \textcolor{black}{the processing of high sequence number packets within a stream is restricted only by the low sequence number packets belonging to the same stream. Thus,} HOL blocking problem is reduced to a stream rather than the entire association.}

\textcolor{black}{Having discussed the fundamentals, we next explore SCTP's support for handover scenarios, real-time traffic, and investigate if the protocol can provide QoS. First things first; let us briefly speak of a very interesting variant of SCTP, i.e., dynamic address reconfiguration (DAR-SCTP) \cite{stewart2003stream, stewart2005stream, maruyama2007stream}}. \textcolor{black}{DAR-SCTP provides the ability to dynamically add or remove an IP address from an association. \textcolor{black}{SCTP} combined with DAR is referred to as mobile SCTP (m-SCTP) \cite{koh2005mobile}. m-SCTP in particular is designed to enable hand-offs in mobile environments by being able to add, delete or  change an IP address. However, it does not support multihoming. Budzisz et al. further studied the applicability of concurrent multipath transfer to handover scenario by examining the distribution of data between two paths of an mSCTP association and proposed mSCTP-CMT \cite{budzisz2009concurrent}.}

\textcolor{black}{SCTP inherently provides reliable data delivery service. However, in order to support real-time applications, partially reliable SCTP (PR-SCTP) \cite{stewart2002sctp} was put forward.} \textcolor{black}{Furthermore, in order to make SCTP quality of service (QoS) aware, wireless multipath multiflow SCTP (WM$^2$-SCTP) \cite{yuan2010extension} has been proposed that groups streams into subflows depending upon QoS requirement.}



\textcolor{black}{With this, we come to the end of our discussion on SCTP. So far, we have covered SCTP basics (in terms of connection setup, flow control, sequence number splitting, ACK, and flow scheduling mechanisms), HOL blocking \& unnecessary fast retransmission problem, and SCTP's support for handovers, real-time traffic \& QoS.} The extensive literature on SCTP is a proof of the immense attention that it received from the research community for about a decade.  \textcolor{black}{Interested reader can further refer to \cite{fu2004sctp, budzisz2012taxonomy}} for a more through study on SCTP. Unfortunately, SCTP \textcolor{black}{has not been} widely adopted due to the lack of support from middleboxes and an API that is distinct from the defacto socket API \cite{paasch2014multipath}. Despite this apparent failure, SCTP established a firm ground for future research in multipathing.

Researchers being convinced of the inevitability of multipathing for future architectures aggressively worked towards more viable solutions, which resulted in the development of MPTCP. \textcolor{black}{The next subsection discusses MPTCP and its variants}.


 


\subsection{Multipath TCP (MPTCP)}
\label{MPTCP}
The standardization of MPTCP \cite{ford2013tcp} is the second major milestone towards \textcolor{black}{the incorporation of the} multipathing trend in the Internet. MPTCP is compatible with the current network infrastructure as it supports middleboxes by either using its own design parameters or seamlessly falling back to regular TCP in case of a conflict.  \footnote{However, there are some cases when MPTCP is problematic with middleboxes. Section \ref{Interworking with middleboxes} overviews this issue.} MPTCP is also \textcolor{black}{backward compatible with existing applications (e.g., MPTCP supports the socket API, which is the defacto networking standard).} These two features promise the success of MPTCP in today\textcolor{black}{'}s and future architectures. 

\textcolor{black}{MPTCP is based on the principle of spreading traffic over multiple network interfaces and balancing load in the network. An MPTCP connection can establish one or more subflows, each of which appears like a regular TCP connection to the network (as indicated by Fig. \ref{fig9}).} \textcolor{black}{Let us now discuss the meat and potatoes of MPTCP in the same pattern as we did for SCTP, i.e., connection establishment, flow control, sequence number splitting, ACK, and flow scheduling mechanisms.}

\textcolor{black}{Connection is established in MPTCP using a 3-way handshake, just like TCP \cite{raiciu2012hard}. In particular, MP\_CAPABLE option in the options field of TCP header identifies that the communication protocol used by end-points is MPTCP rather than TCP.} \textcolor{black}{In order to add subflows to a connection, MP\_JOIN is used.}
Further, a unique token is associated with each connection and additional subflows are added to a particular connection using that unique token \cite{paasch2014multipath}. \textcolor{black}{Reader can consult Fig. \ref{Connection Establishment} for a pictorial representation of connection setup in MPTCP.}

\begin{table*}[!ht]
\scriptsize
\centering
\caption{Comparison of TCP, UDP, SCTP, and MPTCP transport protocols}
\label{tab:comparison}
\begin{tabular}{p{1.5cm}p{1cm}p{1cm}p{1.25cm}p{1.5cm}p{1cm}p{1.3cm}p{1.3cm}p{1cm}p{1cm}p{1.6cm}}

\hline
\cellcolor[HTML]{EFEFEF}\textbf{\emph{Protocol}}
& \cellcolor[HTML]{EFEFEF}\textbf{\emph{Connection Oriented}}
& \cellcolor[HTML]{EFEFEF}\textbf{\emph{Reliability}}
& \cellcolor[HTML]{EFEFEF}\textbf{\emph{Multihoming}}
& \cellcolor[HTML]{EFEFEF}\textbf{\emph{Multistreaming/ Multiple Subflows}}
&
\cellcolor[HTML]{EFEFEF}\textbf{\emph{Ordered Delivery}}
&
\cellcolor[HTML]{EFEFEF}\textbf{\emph{Keepalive Heartbeat Messages}}
&
\cellcolor[HTML]{EFEFEF}\textbf{\emph{Path MTU Support}}
&
\cellcolor[HTML]{EFEFEF}\textbf{\emph{Unordered Delivery}}
&
\cellcolor[HTML]{EFEFEF}\textbf{\emph{Message Oriented}}
&
\cellcolor[HTML]{EFEFEF}\textbf{\emph{Partially Reliable Data Transfer}}
\\
\hline
UDP \cite{postel1980rfc} & $\times$ & $\times$ & $\times$ & $\times$ & $\times$ & 
$\checkmark$ & $\times$ & $\checkmark$ & $\checkmark$ & $\times$ \\
TCP \cite{postel1981transmission} & \checkmark & $\checkmark$ & $\times$ & $\times$ & $\checkmark$ & 
$\checkmark$ & $\checkmark$ & $\times$ & $\times$ & $\times$ \\
SCTP \cite{stewart2000stream} & $\checkmark$ & $\checkmark$ & $\checkmark$ & $\checkmark$ & $\checkmark$ & 
$\checkmark$ & $\checkmark$ & $\checkmark$ & $\checkmark$ & $\checkmark$ \\
MPTCP \cite{ford2011architectural} & $\checkmark$ & $\checkmark$ & $\checkmark$ & $\checkmark$ & $\checkmark$ & 
$\checkmark$ & $\checkmark$ & $\times$ & $\times$ & $\checkmark$ \\
\hline
\end{tabular}
\end{table*}

\textcolor{black}{As mentioned before, flow control can be performed on a per connection or per subflow basis.} However, per subflow control may lead to a deadlock. In order to understand this, consider an example. Suppose there are two subflows (or paths) established by an MPTCP connection. One of the paths stalls due to an outage and at same time the receiver buffer corresponding to the second path is filled to its capacity. In this case, packets from path 2 can not be sent to the application because packets from path 1 are missing. Also, there is no space in the window of path 2 to resend packets from path 1 that are missing. In order to avoid such situations, \textcolor{black}{MPTCP uses per connection flow control}, whereby a shared buffer is used for flow control over all paths. 

MPTCP uses two separate sequence spaces; one per-subflow called subflow sequence number (SSN) to detect losses and second per connection known as data sequence number (DSN) for reconstruction of original message at the
receiver. The subflow level data is mapped at connection level through \emph{data sequence mapping} to retrieve original message.

MPTCP uses ACKs at both connection level as well as at subflow level in order to provide reliable service to users. MPTCP uses SACKs or cumulative ACKs at subflow level while DSN-ACKs are used at connection level for acknowledging received segment. 

\textcolor{black}{In order to take advantage of bandwidth aggregation and resiliency that is provided by MPTCP, it is important to take into account path characteristics. This is due to the fact that diverse path characteristics can lead to unordered data delivery at the receiver because packets travel along paths with varying loss and delay properties. This} unordered data delivery at receiver can lead to unnecessary fast retransmissions that waste useful bandwidth, HOL blocking problem, and undesirable reduction in the size of congestion window. 

\textcolor{black}{\textcolor{black}{For the lessening of packet reordering problem,} some scheduler based solutions that transmit data out of order at sender to ensure in order delivery at the receiver have been proposed in literature.  In particular, Yang and Amer \cite{yang2014using} proposed an MPTCP scheduler based on one-way
communication delay. Le and Bui \cite{le2015forward}\textcolor{black}{,} in agreement with the aforementioned concept, implemented \textcolor{black}{forward-delay-based} packet scheduling (FDPS) algorithm that transmits packets across multiple paths based on delay and bandwidth estimation of various paths in the forward direction (i.e., sender to receiver).
}

\textcolor{black}{HOL blocking is a serious problem and aggravates in diverse network conditions.} While SCTP is capable of handling this problem by using the concept of streams, MPTCP has directly inherited this mess from TCP. Li et al. attempted to compensate for the undesirable effects of HOL blocking problem by introducing network coding to subflows (NC-MPTCP) \cite{li2012network}. \textcolor{black}{NC-MPTCP avoids retransmissions in case of delayed or HOL segments by utilizing redundant data. Systematic coding MPTCP (SC-MPTCP) \cite{li2013tolerating} uses redundant coded packets to alleviate the problem of unnecessary fast retransmissions while also minimizing encoding/ decoding operation. Cui et al. \cite{cui2015fmtcp} further proposed \textcolor{black}{fountain-code-based} multipath TCP (FMTCP) for the same purpose.
Zhou and Shi realized that the core reason for unordered data delivery is the disparity in the end-to-end delay observed over multiple paths \cite{zhou2013goodput}. This observation provided the basis for congestion window adaptation MPTCP (CWA-MPTCP) \cite{zhou2013goodput}, which adjusts the congestion window of each subflow in a way to maintain approximately same end-to-end delays over multiple paths}.

\textcolor{black}{After studying the essentials, let us look into the matter of MPTCP's support for handovers, real-time traffic, and ability to provide QoS. The feasibility of MPTCP for mobile/ Wi-Fi handover has been studied by Paasch et al. \cite{paasch2012exploring}. Specifically, the use of multiple interfaces in smart phones provides better throughput but at the cost of high energy consumption. Like DAR-SCTP, MPTCP can also add or remove an IP address from the connection. Paasch et al. \cite{paasch2012exploring} studied and evaluated three handover modes; \emph{full-MPTCP}, \emph{backup}, and \emph{single-path} in their paper by considering different user requirements (e.g., high throughput, battery lifetime, and traffic pricing).}

MPTCP was originally designed to be fully reliable and fully ordered, this makes it unsuitable for real-time traffic. Realizing this, Diop et al. \cite{diop2012qos} proposed QoS-MPTCP that has
the option of partial reliability and is capable of supporting real
applications (e.g., interactive video applications). 

\textcolor{black}{Let us wrap up our discussion on MPTCP with a quick summary. Just like SCTP, we began our discussion with MPTCP preliminaries (e.g., connection setup, flow control, sequence number splitting, ACK, and flow scheduling mechanisms). Subsequently, we  moved towards the problem of unnecessary fast retransmissions \& HOL blocking and studied how MPTCP variants work around them. Lastly,  MPTCP's support for handovers and real-time traffic was discussed.}

Having discussed the two most famous efforts towards transport-layer multipathing, we conclude this subsection with a table that summarizes the particulars of the well known single and multipath protocols (i.e., UDP, TCP, SCTP, and MPTCP). Table \ref{tab:comparison} intends to give reader a quick overview of the protocols' specifications. Next, we describe some other noticeable past efforts towards multipath transport layer.




\subsection{Other Noticeable Attempts}
\textcolor{black}{This subsection highlights the key features of some known past attempts towards multipath transport-layer protocols.} \textcolor{black}{Like SCTP and MPTCP, the discussion in this section follow suit.} 

\textcolor{black}{The various multipath \textcolor{black}{transport-layer} protocols proposed in literature are either based on TCP or SCTP. As previously discussed, TCP uses a 3-way handshake while SCTP uses a 4-way handshake. Consequently, the multipath transport protocols based on TCP use \textcolor{black}{a} 3-way handshake for establishing connection while those based on SCTP setup connection via 4-way handshake. The reader can have a quick look on the connection setup phase of several famous protocols by referring to Table \ref{tab:SCTP and its variants}.}

\textcolor{black}{Multipath \textcolor{black}{transport-layer} protocols have the liberty of implementing flow control on either \textcolor{black}{a} per connection or \textcolor{black}{a} per path basis, depending on their requirements. R-MTP \cite{magalhaes2001transport} is an exception to this rule, it uses a predefined maximum rate (that determines the number of packets sent per second) with per connection flow control. The reader can further turn to  Table \ref{tab:SCTP and its variants} for a glance at the flow control mechanism of various protocols.}

\textcolor{black}{The heterogeneous multipath protocols also enjoy the freedom of using single or double sequence space, based on their preferences. In particular, \textcolor{black}{parallel TCP (pTCP) \cite{hsieh2002ptcp} that consists of  a striped connection manager (SM) for striping  data across multiple TCP-virtual (TCP-v) uses two separate sequence spaces for detecting losses and for data reconstruction. Radial reception control protocol (R$^2$CP) \cite{hsieh2003receiver,kim2005receiver} that is based on reception control protocol (RCP)   \cite{hsieh2003receiver,kim2005receiver}, is a receiver centric approach that also uses two separate sequence spaces. For seeking more information on sequence number splitting mechanism of other protocols, we refer the reader to Table \ref{tab:SCTP and its variants}.}}

\textcolor{black}{A diverse range of ACK mechanisms aimed at achieving different goals (e.g., reducing retransmissions, taking into account the asymmetry between forward and reverse paths) have been proposed in literature.} \textcolor{black}{\textcolor{black}{In particular,} Lee et al. \cite{lee2002improving} suggested the use of delayed ACK schemes at the receiver for solving the problem of fast retransmissions. Chen et al. \cite{chen2004multipath} proposed multipath TCP (M-TCP), in which they  suggested duplicating each TCP packet and transmitting a copy of it over multiple paths to cater for lossy and mobile environments. Rojviboonchai and Aida \cite{rojviboonchai2004evaluation} pointed out that loss of ACKs on reverse path may result in under utilization of forward path as congestion control mechanism is based on counting ACKs. Thus, robust ACK schemes that take into account the asymmetry between the forward \& reverse paths by transmitting  multiple copies of ACK through multiple paths were presented.} \textcolor{black}{Rate-based M/TCP (R-M/TCP) \cite{rojviboonchai2005rm}, an extension to M/TCP, improves reliability and performance by maintaining \emph{AckedInfoList} for each path to keep information of ACKed data packets. Concurrent TCP (cTCP) \cite{dong2007concurrency} 
\textcolor{black}{consists of} an ACK processor that maintains a list of all packets that have been transmitted but not acknowledged and thus solves gap report problem.
\textcolor{black}{Table \ref{tab:SCTP and its variants} sums up the ACK mechanism of the aforesaid protocols.}}


\textcolor{black}{Moving on to the flow scheduling algorithms, \textcolor{black}{Cetinkaya and Knightly  \cite{cetinkaya2004opportunistic} recommended opportunistic multipath scheduling (OMS) to opportunistically take advantage of good paths (i.e., paths with low loss rates and delay). \textcolor{black}{Dillip Sarkar \cite{sarkar2006concurrent} proposed concurrent multipath TCP (cmpTCP) that schedules packets from a common transmission queue on multiple paths in a WRR fashion. Same principle, as observed in cmpTCP, works for cmpTCPW \cite{sarkar2006qrp04}}}. Once again, we request the reader to consult Table \ref{tab:SCTP and its variants} for a brief look over the flow scheduling mechanisms that are in use by a large number of multipath transport protocols.}

\textcolor{black}{In order to cater for the problem of unnecessary fast retransmitions, cmpTCP and cmpTCPW maintain a retransmission queue per path. Multipath loss tolerant (MPLOT) \cite{sharma2008mplot, sharma2012transport} protocol on the contrary, makes clever use of erasure codes to provide protection against packet losses and decouples in order delivery from reliability. Zhang et al. proposed mTCP \cite{zhang2004transport} that separates the decision of when to send a packet (i.e., congestion control), which packet to send (i.e., reordering), and which paths to use. For the purpose of saving space, we do not delve into the details of how the HOL blocking and the unnecessary fast retransmission problem is tackled by miscellaneous protocols. Reader can seek further information by referring to the respective papers that are indicated in Table \ref{tab:SCTP and its variants}.} 


\textcolor{black}{Following that, we briefly explore the protocols that provide handover capability and can handle real-time traffic. In particular, R$^2$CP \cite{hsieh2003receiver,kim2005receiver} and pTCP \cite{hsieh2002ptcp} are capable of supporting handovers.} \textcolor{black}{
Generally speaking, \textcolor{black}{transport-layer protocols} (except PR-SCTP and QoS-MPTCP) need additional support from application layer in order to support real-time traffic.}
\textcolor{black}{In keeping with this thought, multiflow RTP (MRTP) \cite{mao2006mrtp} and multipath RTP (MPRTP) \cite{singh2013mprtp}, two multipath efforts based on RTP have been proposed. MRTP partitions real-time data across multiple flows while MPRTP splits a single RTP into constant bit rate streams across multiple paths.} \textcolor{black}{Both MRTP and MPRTP are implemented at the application layer. A lot of literature exists on \textcolor{black}{application-layer} multipathing, but as our paper mainly focuses on transport layer so we will not go into any further details of multipath support at the application layer. \textcolor{black}{The} interested reader is referred to \cite{singhsurvey}.}



\textcolor{black}{Existing multipath transport protocols suffer from two major limitations--- \textcolor{black}{specifically, they are application specific, and they may require upgradation of network devices (which is extremely costly).} 
Zhang et al. \cite{zhang2014general} proposed a generalized framework for multipath transport system based on \textcolor{black}{application-level} relays (MPTS-AR) to overcome these two restrictions. MPTS-AR works at the application layer, deploys a large number of \textcolor{black}{application-level} relays to allow end points to send data over one or multiple paths (paths include default path and relay paths) in a single session.} 


This winds up our discussion on several well established attempts towards transport-layer multipathing. Up to this point, we have briefly examined various multipath protocols in terms of their \textcolor{black}{fundamentals (i.e., connection setup,} flow control, sequence number splitting, ACK, and flow scheduling mechanism\textcolor{black}{) and their capability to handle unnecessary fast retransmissions, HOL blocking, handovers,  and real-time traffic}. Next, we move towards the \textcolor{black}{principal} feature of transport layer, i.e., congestion control. 
Bearing in mind its importance for path selection, load balancing, and ensuring fairness with other Internet traffic, we dig deep in this area by taking into consideration the mathematical models and specifics of rate control for multiple paths. 

\section{Multipath Congestion Control}
\label{Multipath Congestion Control}
\begin{quote}
\textcolor{black}{\textit{``To improve is to change; to be perfect is to change often."--- Winston Churchill}}
\end{quote}

For multipath rate control, researchers began their study by first extending the NUM framework to multiple paths in order to obtain an appropriate mathematical model for the problem. Afterwards, efforts towards developing various multipath rate control algorithms began that resulted in a number of solutions. Despite the fact that these various solutions are deployable in practice, they suffer from some limitations and multipath rate control continues to be a hot area of research. The mathematical model for rate control and various congestion control algorithms \textcolor{black}{are presented in the subsequent subsections along with a discussion on their pros and cons.}

\subsection{Rate Control}
\textcolor{black}{Let us start our discussion on rate control by first pondering over the question of how many paths out of all available paths should be used for data transfer. While it seems desirable to use maximum number of paths, their concurrent usage may not always be possible. A large number of paths also means large overhead due to the parallel connections. Thus a balanced trade-off between obtaining the benefits from multiple paths and reducing overhead is required. Researchers have investigated this issue and one of the most important contributions is of Mitzenmacher \cite{mitzenmacher2001power}, who showed that maximum benefits of multipathing can be obtained by the use of just two paths. Key et al. \cite{key2011path} further established that it is not necessary to use all the available paths and it suffices to use only a subset of paths.}

\textcolor{black}{Following this, we briefly look over the process of path selection. Paths are selected primarily on congestion basis and the goal is to increase load on less congested paths and decrease load on more congested paths, thereby balancing load in the network. The selected paths can further be scrutinized based on user's requirements  (i.e., QoS, delay).}

\textcolor{black}{Having established some background in number of paths to use and path selection, we now move towards the heart of this subsection, i.e., rate control. Rate control over multiple paths can be done in either coordinated or uncoordinated fashion. In \emph{coordinated control}, rates over individual paths are determined as a function  of the total number of paths while in \emph{uncoordinated control}, rate for each path is determined independently. Coordinated control over multiple paths can lead to better load balancing properties than uncoordinated control. It is important to note that uncoordinated control can be easily implemented in the network while coordinated control requires some modifications to either TCP or at the application layer \textcolor{black}{\cite{tullimas2008multimedia,wang2009multipath}}.}

The theoretical models of coordinated and uncoordinated controller are realized in practice by using a combination of multipath congestion control algorithms and AQM technology, just like we did before for single-path transport layer. Before getting into the details of multipath congestion control algorithms, it is better to first get acquainted with the fluid flow model for multipath rate control that is the topic of next discussion. 

\subsection{Network Model and Stability}
The utility maximization framework for single-path \textcolor{black}{transport layer} (as presented in section \ref{Network Model and Stability}) can be extended to multiple paths. In such a case, the optimization problem simply extends to maximizing the aggregate utility of multihomed users subject to link constraints.
\begin{equation}
\label{Equation11}
\max \sum_{s \in S} U_{s}(\textbf{x}_s) \quad \text{subject to} \quad y_l \leq c_l
\end{equation}
where $\textbf{x}_s= (x_r, r \in s)$. The above equation is motivated as follows: let source $s$ has multiple routes $r=\{r_1,r_2\}$ to destination, $s$ splits its flow across $r_1$ (subflow 1) and $r_2$ (subflow 2) according to a specific multipath \textcolor{black}{congestion} control algorithm. Since $s$ is the unique source associated with these routes, so $s(r_1)=s(r_2)=s$. The two subflows transmitted through $r_1$ and $r_2$ ultimately reach the destination where they combine to reconstruct the original message. It is important to remember that the aggregate transmission rate of various sources that are transmitting data across multiple paths, always lie within link capacity. 

Following in the footsteps of classical \textcolor{black}{NUM framework}, equation (\ref{Equation11}) can be formulated into an appropriate primal, dual or primal dual problem so that each source and link can solve the optimization problem using only local information. Same principles of adjusting the transmission rate according to aggregate link prices and adjusting link prices according to aggregate transmission rate follows here. The transmission rate is adjusted using multipath congestion control algorithms and prices are adjusted using AQM technology.


In order to establish the stability of fluid flow models when multiple paths are used, Han et al. \cite{han2004overlay} used a framework that implemented application-level overlay routers. They discovered that network is stable if the responsiveness of a route serving a source-destination pair is constrained by the RTTs of all routes that are serving the same source-destination pair. This stability condition however, does not cater for different RTTs.

In order to account for heterogeneous RTTs, Kelly and Voice \cite{kelly2005stability} used a fluid flow model to analyze the local stability of an end-to-end joint routing and rate control algorithm. It was found that the network is stable if the responsiveness of each route is constrained by the RTT of only that route (the information of which is readily available at the transport layer). 

While stability and equilibrium properties of  congestion control algorithms can be studied using utility maximization framework, there exist some cases in which \textcolor{black}{the} utility function for a congestion control algorithm does not exist e.g., \textcolor{black}{fully coupled} algorithm. So, how can we study the stability and convergence of such an algorithm? Peng et al. \cite{peng2014multipath} proposed a unified fluid flow model for exploring the stability and equilibrium properties of a broad range of multipath congestion control algorithms. 

\textcolor{black}{This new theoretical framework proposed by Peng et al. \cite{peng2014multipath} not} only generalizes most of the existing congestion control algorithms but also provides a better understanding of the design parameters (in particular TCP-friendliness, responsiveness\textcolor{black}{,} and stability). TCP-friendliness measures how fair the congestion control algorithm is to classical TCP while responsiveness measures how quickly the algorithm responds to changes in network topology. Peng et al. \cite{peng2014multipath}, using this framework discovered an inevitable trade-off between responsiveness and TCP-friendliness. \emph{Balanced linked adaptation} (BALIA), a congestion control algorithm that strikes the most balanced trade-off between TCP-friendliness and responsiveness has been developed on this framework.

Having discussed the fluid flow model and its stability, we now move towards the multipath congestion control algorithms that are responsible for adjusting transmission rates at source hosts. 

\subsection{Congestion Control Algorithms for Multiple Paths}
\label{Congestion Control Algorithms for Multiple Paths}
\textcolor{black}{As already mentioned in Section \ref{Important Question Related to Multipath Transport}, congestion control algorithms can be broadly classified as coupled or uncoupled, both of which are studied in detail in the following paragraphs.}

\textcolor{black}{Uncoupled congestion control maintains a separate congestion window for each path, thereby not implementing resource pooling.} The algorithm can be simply implemented by extending TCP Reno, TCP \textcolor{black}{W}estwood or any other single-path TCP congestion algorithm to multiple paths. While the algorithm is highly responsive to changes in network topology (e.g., loss rate and congestion level), it runs the risk of being unfair to the other Internet traffic in case of shared bottlenecks.

As a first workaround, researchers proposed the idea of identifying and detecting shared bottlenecks and suppressing them, a phenomena called \emph{shared bottleneck detection} (SBD). This can ensure fairness, but the time required to identify and suppress shared bottlenecks can sometimes significantly degrade performance. 

\textcolor{black}{Coupled algorithms, on the contrary, couples the congestion windows of different paths in some proportion, thereby benefiting from resource pooling and addressing the issue of fairness.} \textcolor{black}{However, coupled congestion control algorithms are not as responsive to network changes as uncoupled ones.} We have arranged various multipath \textcolor{black}{transport-layer} protocols in \textcolor{black}{Table \ref{tab:SCTP and its variants}} on the basis of the type of congestion control algorithm used by them.

\textcolor{black}{We have already mentioned in Section \ref{Multipath Transport Layer Protocols} that SCTP, despite providing a solid ground in transport-layer multipathing, has not been widely adopted. MPTCP, on the other hand, has the potential to impact future Internet architectures. That is why we only go into the details of MPTCP's congestion control algorithms.
Reader interested in SCTP's congestion control mechanisms can refer to \cite{dreibholz2011impact, dreibholz2012simulation}}.  

\textcolor{black}{With the dawn of MPTCP, three design objectives for designing congestion control algorithms were identified. These are described below \cite{khalili2013mptcp}:}
\begin{enumerate}
\item A multipath flow should at least perform as well as a single-path flow would on the best  available path.
\item A multipath flow should be fair to other Internet traffic (i.e., it does not take up any more capacity on a path than a single-path flow).
\item A multipath flow should balance congestion in the network by moving traffic from more congested paths to less congested ones (i.e., resource pooling).
\end{enumerate}

Based on the afore-stated goals, various congestion control algorithms have been proposed in literature. With this background, let us delve into the details of \textcolor{black}{window-increase/window-decrease} functions for some famous congestion control algorithms.




\begin{table}
\centering
\scriptsize
\caption{Notations used for Multipath Rate Control Alogorithms}
\label{tab:Notations used in this paper}
\begin{tabular}{p{1.2cm}p{6.3cm}}
\hline
\textbf{\emph{Parameter}} & \textbf{\emph{Specification}} \\ 
\hline
$I_r(\textbf{w}_s)$ & Window increase function for source $s$ on route $r$. \\
$D_r(\textbf{w}_s)$ & Window decrease function for source $s$ on route $r$.\\
$\text{cwnd}_i$ & Size of congestion window maintained at path $i$. \\
$BW_i$ & Bandwidth obtained on path $i$. \\
$\tau_r$ & RTT on route $r$. \\
$x_r$ & Transmission rate on route $r$.\\
$w_r$ & Congestion window size on route $r$. \\
$\sum_{k \in s} w_k$ & Total window size for source $s$. \\
$a$ & Used to control aggressiveness on various paths.\\
$\beta$ & Used to control aggressiveness. \\
$n$ & Number of paths.\\
$N_m$ & It is the number of flows belonging to bottleneck sharing group m.\\
$N_c$ & It is the total number of MPTCP connections. \\
$\gamma_r$ & Uses both set of flows that are using best paths in terms of throughput and set of flows that have largest congestion window as parameters. This parameter balances load in the network by:
\begin{enumerate}
\item Increasing traffic on best paths, i.e, $\gamma$ is positive.
\item Decreasing traffic on congested paths, i.e $\gamma$ is negative.
\item Do nothing, i.e., $\gamma$ is zero. 
\end{enumerate} \\
$\gamma_r'$ & Only uses set of flows that are using best paths as parameter.\\
$\lambda_r$ & It is equal to $\frac{max \{x_k\}}{x_r}$.\\
\hline
\end{tabular}
\end{table}

\textcolor{black}{Multipath congestion control algorithms, like single-path algorithms, are designed around the principle of additive increase (AI) in window size in case of a successful transmission and multiplicative decrease (MD) when a transmission failure occurs. This AIMD approach consists of four algorithms namely; slow start, congestion avoidance, fast retransmit\textcolor{black}{,} and fast recovery. Our discussion is focused only on the congestion avoidance phase of various congestion control algorithms. The increase and decrease in window size is denoted by $I_r(\textbf{w}_s)$ and $D_r(\textbf{w}_s)$ respectively.} The aggregate price of a route $q_r$ is directly proportional to the probability of a transmission failure, since high $q_r$ indicates very little probability of successful transmission. Taking $q_r$ as the probability of transmission failure, the probability of a successful transmission is $1-q_r$. The change in the size of congestion window ($\Delta w_r$) can thus be represented as:
\begin{equation}
\label{Window Increase Decrease function}
\Delta w_r = \{I_r(\textbf{w}_s)(1-q_r)-D_r(\textbf{w}_s)q_r\} w_r
\end{equation}

This model (\ref{Window Increase Decrease function}) has been adapted from \cite{peng2014multipath}. The multipath congestion control algorithms increase $I_r(\textbf{w}_s)$ the window size of a source $s$ ($\textbf{w}_s$) on a route $r$ if the transmission is successful otherwise reduce it using the specific decrease function, i.e.,  $D_r(\textbf{w}_s)$. The amount by which the window size is increased or decreased is specific to each protocol. \textcolor{black}{The window-increase/window-decrease function of various multipath congestion control algorithms together with their strengths are weaknesses} are summarized in Table \ref{tab:CongestionControlAlgos}. \textcolor{black}{The} parameters used in the \textcolor{black}{window-increase/window-decrease} function are explained in Table \ref{tab:Notations used in this paper}. An important thing to note here is that \emph{fully coupled}, \emph{semi-coupled}, \emph{linked increase adaptation} (LIA), \emph{opportunistic linked increase adaptation} (OLIA), \emph{dynamic window coupling (DWC)}, \emph{adapted OLIA} (AOLIA)\textcolor{black}{,} and BALIA are all examples of coupled algorithms.  Next, we discuss the \textcolor{black}{congestion} control algorithms in detail and compare their performance in terms of TCP-friendliness and responsiveness. 

\begin{figure} 
\centering
\includegraphics[width=8cm]{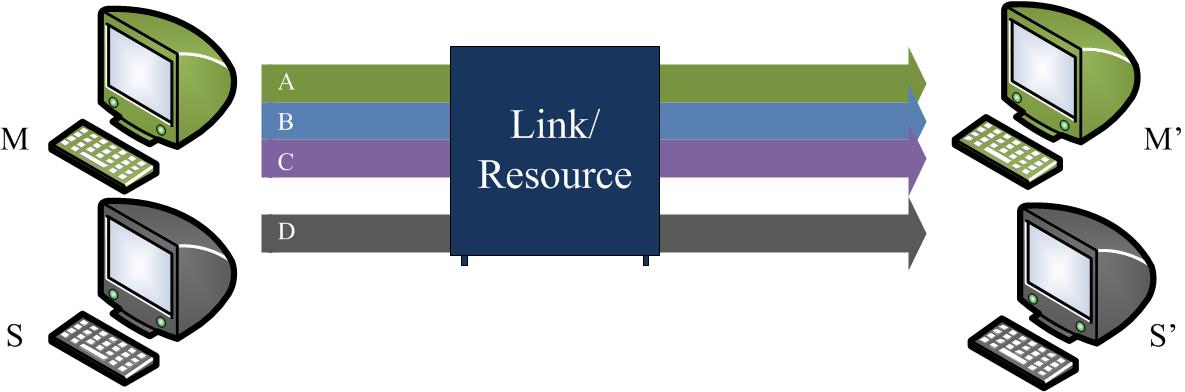}
\caption{Example case}
\label{fig13}
\end{figure}

\begin{table*}[!ht]
\centering
\scriptsize
\caption{Congestion Control Algorithms for Multipath Transport}
\label{tab:CongestionControlAlgos}
\begin{tabular}{p{1.5cm}p{.5cm}p{3.7cm}p{1.5cm}p{0.5cm}p{0.9cm}p{0.5cm}p{0.8cm}p{1.1cm}p{1.6cm}p{1cm}}
\hline

\cellcolor[HTML]{EFEFEF} \textbf{\emph{Algorithms}} & \cellcolor[HTML]{EFEFEF}\textbf{\emph{Year}} & 
\cellcolor[HTML]{EFEFEF}\textbf{\emph{$I_r(\textbf{w}_s$)}} &
\cellcolor[HTML]{EFEFEF}\textbf{\emph{$D_r(\textbf{w}_s$)}} &
\multicolumn{7}{c}{\cellcolor[HTML]{EFEFEF} \textbf{\emph{Strengths and Weaknessess}}} \\

\cellcolor[HTML]{EFEFEF} &
\cellcolor[HTML]{EFEFEF}&
\cellcolor[HTML]{EFEFEF}&
\cellcolor[HTML]{EFEFEF}& 
\cellcolor[HTML]{EFEFEF}\textbf{\emph{Flappy}} & 
\cellcolor[HTML]{EFEFEF}\textbf{\emph{Capture Problem}} & 
\cellcolor[HTML]{EFEFEF}\textbf{\emph{Diverse RTTs}}& 
\cellcolor[HTML]{EFEFEF}\textbf{\emph{Resource Pooling}} & 
\cellcolor[HTML]{EFEFEF}\textbf{\emph{Efficient Network Utilization}} & 
\cellcolor[HTML]{EFEFEF}\textbf{\emph{Distinguishes b/w shared \& distinct bottlenecks}} & 
\cellcolor[HTML]{EFEFEF}\textbf{\emph{Incast Collision}}
\\ 
\hline

Uncoupled Control & 2000s  & $\frac{1}{w_r}$  & $\frac{w_r}{2}$ & $\times$ & $\times$ & $\checkmark$ & $\times$ & $\times$ & $\times$ & $\times$
\\

EW-TCP \cite{honda2009multipath} & 2009 & $\frac{1}{w_r \sqrt{n}}$ & $\frac{w_r}{2}$ & $\times$ & $\times$ & $\times$ & $\checkmark$ & $\times$ & $\times$ & $\times$\\


Fully Coupled \cite{kelly2005stability, han2006multi, wischik2011design} & 2011 & $\frac{1}{\sum_{k \in s} w_k}$ & $\frac{\sum_{k \in s} w_k}{2}$ & $\checkmark$ & $\checkmark$ & $\times$ & $\checkmark$ & $\checkmark$ & $\times$ & $\times$
\\

Semi-coupled \cite{wischik2011design} & 2011   & $\frac{a}{\sum_{k \in s} w_k}$ & $\frac{w_r}{2}$  & $\times$ & $\times$ & $\times$ & $\checkmark$ & $\checkmark$ & $\times$ & $\times$
\\

LIA (RFC 6356) \cite{raiciu2011coupled} & 2011 & min $(\frac{\text{max}_r (w_r/\tau_r^2)}{(\sum_{k \in s} w_k/ \tau_k)^2}, \frac{1}{w_r})$ & $\frac{w_r}{2}$ & $\times$ & $\times$ & $\checkmark$ & $\checkmark$ & $\checkmark$ & $\times$ & $\times$
\\

DWC \cite{hassayoun2011dynamic} & 2011 & min $(\frac{ \tau_r}{N_m \text{min}_{r \in m} \tau_r  \sum_{k \in s} w_k}, \frac{1}{w_r} )$ & $\frac{w_r}{2}$ & $\times$ & $\times$ & $\checkmark$ & $\checkmark$ & $\checkmark$ & $\checkmark$ & $\times$
\\

OLIA \cite{khalili2012non,khalili2013mptcp} & 2012$,$ 2013  & $\frac{w_r/ \tau_r^2}{(\sum_{k \in s} w_k/ \tau_k)^2} + \frac{\gamma_r}{w_r}$ & $\frac{w_r}{2}$ & $\times$ & $\times$ & $\checkmark$ & $\checkmark$ & $\checkmark$ & $\times$ & $\times$
\\

AOLIA \cite{singh2013enhancing} & 2013 & min$(\frac{\text{max}_r (w_r/\tau_r^2)}{(\sum_{k \in s} w_k/ \tau_k)^2} + \frac{\gamma_r'}{w_r},\frac{1}{w_r})$  & $\frac{w_r}{2}$ & $\times$ & $\times$ & $\checkmark$ & $\checkmark$ & $\checkmark$ & $\times$ & $\times$
\\

BALIA \cite{PengWL13, peng2014multipath} & 2013$,$ 2014 & $ \frac{x_r}{\tau_r (\sum_{k \in s} x_k)^2} (\frac{1+\lambda_r}{2})( \frac{4+\lambda_r}{5}) $ & $\frac{w_r}{2} \text{min} ( {\lambda_r,1.5})$ & $\times$ & $\times$ & $\checkmark$ & $\checkmark$ & $\checkmark$ & $\times$ & $\times$ 
\\ 

EW-MPTCP \cite{ming2014mptcp} & 2014 & min$  (\frac{\frac{\beta}{\sum_{k \in s} w_k},\frac{1}{w_r}}{N_c})$ & $\frac{w_r}{2}$ & $\times$ & $\times$ & $\checkmark$ & $\checkmark$ & $\checkmark$ & $\times$ & $\checkmark$
\\

\hline
\end{tabular}
\end{table*}

We start our discussion with Fig \ref{fig13}. The figure shows a multihomed pair $MM'$ and a single homed pair $SS'$ that are sharing a bottleneck. $MM'$ has three paths $A$, $B$\textcolor{black}{,} and $C$ while $SS'$ has a single-path $D$. The available bandwidth, RTT\textcolor{black}{,} and the size of congestion window that a user has on path $i$ is represented by BW$_i$, RTT$_i$\textcolor{black}{,}  and cwnd$_i$ respectively. The subsequent equations represent the characteristics of these four paths.
\begin{equation}
\text{BW}_A \approx \text{BW}_D \geq \text{BW}_B \geq \text{BW}_C
\end{equation}
\begin{equation}
\text{RTT}_A \approx \text{RTT}_D \leq \textcolor{black}{\text{RTT}}_B \leq \textcolor{black}{\text{RTT}}_C
\end{equation}
\begin{equation}
\text{cwnd}_A \approx \text{cwnd}_D \leq \text{cwnd}_B \leq \text{cwnd}_C
\end{equation} 

It is clear from these equations that $A$ is the best path and $C$ is the worst path. All the congestion control algorithms described below are studied using this example model. 

Let us first consider the naive approach of extending traditional TCP's  congestion control mechanism to multiple paths. When applied to our example case: $MM'$ obtains a bandwidth of $\text{BW}_A+\text{BW}_B+\text{BW}_C$ while $SS'$ gets $\text{BW}_D$. Clearly, this algorithm is unfair to TCP flows. On the bright side, the algorithm is very responsive to changes in network topology, i.e., it responds to RTT and losses on individual paths by adjusting the size of congestion window on each path accordingly. As previously mentioned, the initial fixes towards solving fairness issue was SBD, which might take some time in identifying and suppressing subflows traversing shared bottlenecks.

\textcolor{black}{Realizing the fact that SBD may be slow in responding to shared bottlenecks, evenly weighted TCP (EW-TCP) \cite{wischik2011design} was proposed.
EW-TCP evenly splits a flow among all the available paths and does not require SBD. Applying EW-TCP to Fig. \ref{fig13},} $MM'$ obtains an aggregate bandwidth of $\frac{\text{BW}_A+\text{BW}_B+\text{BW}_C}{3}$ (Mbps). Thus, the algorithm is less aggressive towards TCP flows in comparison to uncoupled Reno. EW-TCP equally splits flow irrespective of path characteristics, consequently it is unable to efficiently use the network (e.g., a better  utilization would have been to send most traffic on $A$ as it is the best path and least on $C$). By not taking path characteristics into account, EW-TCP becomes problematic when paths have heterogeneous RTTs.

Moreover, evenly splitting a link bandwidth among multiple subflows may not conform to design goal 1 (i.e., a multipath flow should at least perform as well as a single-path flow). In such cases, weighted splitting can be performed. Honda et al. \cite{honda2009multipath} essentially based their work on EW-TCP but with using different weights on each path and adapting the weights in order to achieve fairness and better network utilization. However, their work could not handle heterogeneous RTTs and adaptive weighted TCP (AWTCP) \cite{hu2011research} was proposed to overcome the RTT limitation of EW-TCP. 

The beautifully simple concept of sending more traffic on paths with better path characteristics (i.e., low drop probability, small delay, small size of congestion window) to balance load and reduce congestion in the network was originally presented by Kelly and Voice \cite{kelly2005stability}. Later on, Han et al. \cite{han2004overlay} worked in the same direction to improve the differential equation models proposed by Kelly and Voice \cite{kelly2005stability}. Based on these two works, fully coupled algorithm that couples the window-increase and window-decrease function of congestion windows across all subflows was presented. 

Applying fully coupled algorithm to our example scenario, most traffic will be sent on path $A$ and least on path $C$. In this case, the aggregate throughput obtained by $MM$' will be such that:
\begin{equation}
\text{BW}_A + \text{BW}_B + \text{BW}_C \approx \text{BW}_D
\end{equation}

Thus, fully coupled effectively utilizes the network and is fair to TCP flows due to the weighted splitting. However, it is problematic due to capture problem, flappiness, unnecessary reduction in the size of congestion window\textcolor{black}{,} and heterogeneous RTTs. 

In order to understand flappiness, let us assume that $A$ and $B$ have approximately similar path characteristics (i.e., $\text{BW}_{A} \approx \text{BW}_B$, $\text{RTT}_{A} \approx \text{RTT}_B$\textcolor{black}{,} and $\text{cwnd}_A \approx \text{cwnd}_B$). Both these paths are vulnerable to small random fluctuations (i.e., small changes in bandwidth, congestion window size), a phenomena that is not rare in practical networks. The response of fully coupled algorithm to these random fluctuations is flipping the subflow between two optimal paths (i.e., $A$ and $B$), since it treats these temporary variations as permanent changes. This makes the algorithm unstable.  

In our example scenario, $C$ has the least traffic because it is the worst path. This can eventually lead to a state where all traffic is transmitted via $A$ \& $B$ and $C$ is not used at all. Later in time, if the path characteristics of $A$ and $B$ worsen and that of $C$ improves, we cannot utilize $C$ as there is not sufficient probing traffic on it. The flow in such a case is said to be ``trapped'' or ``captured'' in the less desirable paths and can not utilize the available better paths. Additionally, fully coupled algorithm does not account for heterogeneous RTTs. Since it fully couples the window decrease function, a single bad path can unnecessarily reduce the congestion window of good paths too.

\textcolor{black}{Wischik et al., \cite{wischik2010balancing}} while trying to solve flappiness and capture problem, proposed the principle of \textcolor{black}{\textit{equipoise}}. \textcolor{black}{Equipoise defines a} trade-off between resource pooling and traffic balancing by stating that congestion control algorithms should balance traffic among \textcolor{black}{the} best paths in a way such that it is robust to transient network fluctuations \textcolor{black}{while also being} responsive to persistent changes. This principle influenced all the subsequently proposed algorithms in literature. 

Semi-coupled algorithm \cite{wischik2011design} was designed in line with the notion of equipoise. It improved fully coupled by solving capture problem, flappiness\textcolor{black}{,} and decoupling window-decrease function of multiple paths. The algorithm performs weighted splitting (with a bias towards less congested paths) but keeps a small amount of traffic on all paths. It reaps the benefits of resource pooling by coupling only the increase function while window-decrease functions are decoupled from each other. This decoupling reduces unnecessary reduction in the size of congestion window on good paths due to bad ones. However, the algorithm does not cater for heterogeneous RTTs. 

LIA \cite{raiciu2011coupled} aims to improve \textcolor{black}{the} semi-coupled algorithm by taking into consideration heterogeneous RTTs. The window-increase function of LIA has two parts (refer to Table \ref{tab:CongestionControlAlgos}); \textcolor{black}{the} first part takes into account the size of congestion window and RTT during flow splitting\textcolor{black}{,} while \textcolor{black}{the} second part increases window  size in accordance with classical TCP. The overall window is adjusted by a value that is the minimum of the aforementioned parts, thereby ensuring that aggressiveness of multipath flow is never more than that of single-path TCP. In terms of our example scenario, a multipath flow is split between $A$, $B$\textcolor{black}{,} and $C$ in accordance with their path characteristics. 


LIA has a serious disadvantage that it cannot  differentiate between subflows sharing a common or distinct bottleneck. If the subflows are always assigned a bandwidth assuming that they traverse distinct bottleneck, then the mechanism is obviously unfair to other Internet traffic. On the other hand, if subflows are assigned bandwidth considering a shared bottleneck, then available bandwidth due to subflows traversing distinct bottlenecks can not be utilized. Thus, there is a need to distinguish subflows that traverse a shared bottleneck from ones that use distinct bottlenecks.

DWC \cite{hassayoun2011dynamic} came as a solution to this problem, it couples only those subflows that have a common bottleneck while the congestion control for subflows belonging to distinct bottlenecks is separate. DWC is TCP-friendly, responsive to shifting bottlenecks in the network\textcolor{black}{,} and maximizes throughput over disjoint bottlenecks. This is achieved by classifying subflows into subflow-sets. Each subflow-set represents a distinct bottleneck, congestion windows across subflow-sets are considered independent to achieve maximum throughput. DWC has a centralized entity, called the \emph{subflow manager} that creates and manages \textcolor{black}{window-increase/window-decrease} functions of subflows. 

Khalili et al. \cite{khalili2013mptcp} found out two more limitations of LIA. With carefully designed \textcolor{black}{testbeds} for experimentation, Khalili et al. observed that upgrading some TCP users to MPTCP may reduce the throughput of other users while not improving \textcolor{black}{the} throughput of upgraded users. This is a typical symptom of \emph{non-Pareto optimality}. Investigation into the reasons of this behavior revealed that LIA inherently employs a trade-off between resource pooling and responsiveness, i.e., for good responsiveness LIA departs from Pareto optimality. This can also make LIA more aggressive towards other Internet traffic in some cases. 

OLIA came as a solution to the above-mentioned problems of LIA. Its window-increase function presented in Table \ref{tab:CongestionControlAlgos} consists of two parts. The first part is based on the work of Kelly and Voice \cite{kelly2005stability} and ensures Pareto optimality. The second part guarantees responsiveness and non-flappiness by measuring the bits transmitted since last loss, which allows OLIA to more quickly adapt to network changes. OLIA uses throughput and size of congestion window as two parameters for dynamically adjusting flow rate. Let us apply OLIA to our example model. OLIA:
\begin{itemize}
\item increases data rate on path $A$ by making $\gamma$ positive (as it is the best path with smallest cwnd$_A$).
\item keeps data rate constant on $B$ by making $\gamma$ zero (as $B$ has small cwnd$_B$, but is not an optimal path).
\item reduces data rate on path $C$ by making $\gamma$ negative (as $C$ has largest cwnd$_C$ and it is a bad path).
\end{itemize}   

\begin{figure} 
\centering
\includegraphics[width=8cm]{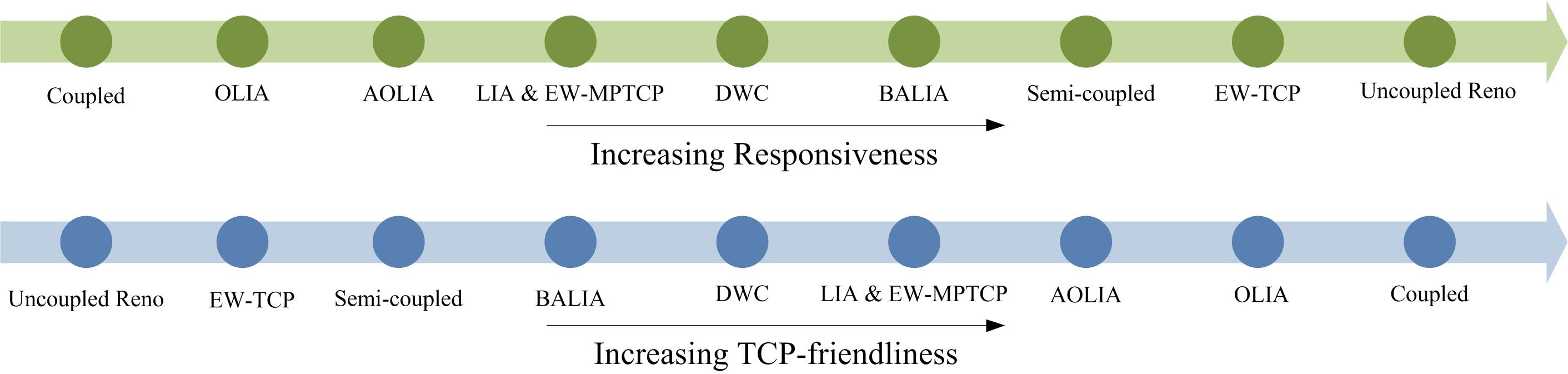}
\caption{Comparison between various rate control algorithms}
\label{fig5}
\end{figure}

Singh et al. \cite{singh2013enhancing} \textcolor{black}{argued} by comparing the first term in window increase function of LIA and OLIA that overall throughput of LIA is usually better than OLIA. In an effort to combine the best of both worlds, Singh et al. proposed adapted OLIA (AOLIA) that is a \textcolor{black}{LIA-based} extension of OLIA. \textcolor{black}{Its window-increase function consists of two parts: the first part is a combination of OLIA and LIA while the second part is based on the regular TCP window-increase function.}

It is important to mention the role of congestion control algorithm in solving \textcolor{black}{the} incast collision problem that frequently occurs in datacenters. Li et al. \cite{ming2014mptcp} are among the pioneers to address this problem. In order to understand \textcolor{black}{the} incast collision, consider a scenario where a receiver issues a request to multiple senders (i.e., one to many communications). The response from all senders to a single receiver can result in creating a hotspot. In particular, top of rack (TOR) switch in datacenter becomes the bottleneck when multiple servers try to respond to a single receiver.
Because MPTCP is unable to efficiently solve this problem, Li et al. \cite{ming2014mptcp} proposed EW-MPTCP. EW-MPTCP is an enhancement to LIA, it weights the size of congestion window in reverse proportion to the number of paths. It is worth noting that EW-MPTCP caters for subflows from differnt MPTCP connections that may be competing at a shared point of congestion while subflows from same MPTCP connection are dealt with using LIA. 

The TCP-friendliness and responsiveness of the aforementioned algorithms is illustrated graphically in Fig. \ref{fig5}. An inevitable trade-off between TCP-friendliness and responsiveness is evident from the figure, i.e., \textcolor{black}{the} most responsive algorithm is least friendly and vice versa. It can be seen that uncoupled Reno is the least \textcolor{black}{TCP-}friendly algorithm while coupled is the most friendly but least responsive algorithm. BALIA strikes the most balanced trade-off between responsiveness and \textcolor{black}{TCP-}friendliness. The TCP-friendliness and responsiveness of EW-MPTCP is same as LIA. Further research efforts towards designing multipath \textcolor{black}{congestion} control algorithms that strike a more efficient trade-off between responsiveness and \textcolor{black}{TCP-}friendliness are in process.


\textcolor{black}{This completes our discussion on the nine design questions that were motivated in Section \ref{Important Question Related to Multipath Transport}. The gist of our study is reiterated in Table \ref{tab:Summary} for a quick recap. Other than these nine questions, we also looked into the HOL blocking problem, unnecessary fast retransmit, and multipath transport protocols support for real-time traffic and handovers. The issue of how many paths to use and path selection was also investigated. In addition to that, we also enquired into the advantages and disadvantages provided by different congestion control algorithms. Let us now move towards the last chapter of our paper, i.e., open research issues and challenges in the field of transport-layer multipathing.}

\section{Issues and Challenges}
\label{Issues and Challenges}
\begin{quote}
\textcolor{black}{\textit{``To reach a port, we must sail--- sail, not tie at anchor--- sail, not drift."--- Franklin D. Roosevelt}}
\end{quote}

Multipathing is a phenomena that is increasingly being \textcolor{black}{adopted} by datacenters and wireless environments. Multipath transport layer is not a mature technology and several research problems exist in this field, which are receiving considerable attention from the  research community \cite{scharfmultipath, kostopoulos2010towards}. The hot research areas in multipath transport layer can be broadly classified into the following four categories:
  
\begin{enumerate}
\item Internetworking with middleboxes.
\item Congestion control with multiple paths.
\item \textcolor{black}{Cross-layer} multipath interactions.
\item Multipathing and green networking.
\end{enumerate} 

The open research issues and challenges in each of the aforementioned areas are next discussed in detail. 


\begin{table}
\centering
\scriptsize
\caption{Comparison between Single-path and Multipath Transport Layer}
\label{tab:Summary}
\begin{tabular}{p{2.2cm}p{2.3cm}p{3cm}}
\hline
\cellcolor[HTML]{EFEFEF}\textbf{\emph{Key Mechanisms}} 
& \cellcolor[HTML]{EFEFEF}\textbf{\emph{Single-path Transport Layer}} 
& \cellcolor[HTML]{EFEFEF}\textbf{\emph{Multipath Transport Layer}} \\ 
\hline
\textcolor{black}{Connection setup} & \textcolor{black}{3-way handshake} & \textcolor{black}{3-way or 4-way handshake} \\
\\
Flow control & Per connection basis & Per connection or per path basis \\
\\
Sequence space & Single & Single or double or  triple \\
\\
ACK & SACK, Cumulative ACK\textcolor{black}{,} and Delayed ACK &  SACK, Cumulative ACK, Delayed ACK\textcolor{black}{,} and Duplicated \& delayed ACK \\
\\
Flow scheduling & \textcolor{black}{N/A} & WRR, OMS, FPS\textcolor{black}{,} and EDPF \\
\\
NUM framework & $\max \sum_{s \in S} U_s(x_s)$& $\max \sum_{s \in S} U_{s}(\textbf{x}_s)$ \\
\\
Congestion control & \textcolor{black}{Uncoupled} & \textcolor{black}{Uncoupled or coupled}\\
\\
\textcolor{black}{Window-Increase/Window-Decrease} & \textcolor{black}{$\frac{1}{w_r}$/$\frac{w_r}{2}$} & \textcolor{black}{Summarized in Table \ref{tab:CongestionControlAlgos} (For MPTCP).}\\
\\
\textcolor{black}{TCP-Friendliness \& Responsiveness} & \textcolor{black}{N/A \&  Responsive} & \textcolor{black}{Summarized in Fig. \ref{fig5} (For MPTCP).}
\\
\hline
\end{tabular}
\end{table}

\subsection{Internetworking with Middleboxes}
\label{Interworking with middleboxes}
The Internet was originally designed around the ``end-to-end design principle" \cite{saltzer1984end}. End-to-end design principle states that complex functions are performed at end hosts while the network implements simple functions (e.g., forwarding packets). Obeying this principle, the deployment of multiple paths would require modifications only at end hosts while the network can operate as such. 

With the increasing popularity of Internet, researchers felt the need for more sophisticated nodes. This \textcolor{black}{lead} to the development of more advanced middleboxes that essentially disrupt the end-to-end design principle. Examples include firewalls that  provide protection against malicious nodes, network address translators (NATs) that solve the problems associated with insufficient IPv4's address space, and load balancers, \textcolor{black}{just to name a few}. These middleboxes are widely deployed in today's Internet, enterprise and cellular networks \cite{sherry2012making}. Thus, it is critical to consider the role of middleboxes while making any architectural change in the network. In general, the dumb middleboxes once capable of only forwarding packets are now intelligent enough to even modify packet headers.

Despite its importance, most multipath transport protocols (e.g., SCTP) were designed assuming that packets arrive at the  destination unmodified. Due to this idealized assumption, their deployment was not successful. MPTCP was the first multipath transport protocol that was designed while keeping in mind the role of middleboxes. MPTCP handles middleboxes by reverting back to TCP in case of a conflict  \cite{ford2013tcp, raiciu2013recent, paasch2014multipath}. For translating MPTCP packets to TCP, Detal et al. \cite{detal2013multipath} proposed a protocol converter MIMBox.

Although MPTCP is capable of handling most of the middleboxes, some corner cases do exist that can degrade the performance of MPTCP \cite{hesmans2013tcp}. 
Consequently, future research efforts towards better handling middleboxes are required.

\subsection{Congestion control with multiple paths}
The open research problems related to \textcolor{black}{congestion control with single-path} were highlighted by the Internet Research Task Force (IRTF) \cite{papadimitriou2011open}. The challenges associated with single-path congestion control are aggravated with the use of multiple paths. Thus, appropriately addressing these challenges is more important than ever. These key challenges are mentioned below.
\begin{itemize}
\item \textit{Heterogeneity.}
The Internet is composed of a vast variety of heterogeneous links and paths that have diverse bandwidths (from several kilo bits to giga bits) and delay (RTTs ranging from millisecond to a second) characteristics. Moreover, the path characteristics are not constant and change with time and traffic loads. The level of heterogeneity in Internet  is expected to grow in future. The design of congestion control algorithms that deal with this vast range of different technologies in a stable and efficient way is a challenging task. 

\item \textit{Interaction between heterogeneous congestion control algorithms.}
Researchers have mostly studied \textcolor{black}{homogeneous} systems that use same congestion control algorithms. The interaction between various congestion control algorithms is not fully understood and demands attention from research community. 

\item \textit{Stability.}
The modeling of realistic network for stability analysis can be extremely difficult if packet sizes and heterogeneous RTTs are taken into account. The common practice is to model simple cases and study more complex behavior using simulations. However, a mechanism that is found to be stable in simulations may be unstable in reality as simulations make simplifying assumptions. Thus, better models that closely approximate real behavior are required.  

\item \textit{Friendliness vs Responsiveness.}
With multiple paths, there exists an inevitable trade-off between responsiveness and friendliness as already mentioned in section \ref{Congestion Control Algorithms for Multiple Paths}. Future research towards developing algorithms that strike a more efficient trade-off between responsiveness and friendliness is desired.

\item \textit{Need for new tools.}
While single-path congestion control tools could be extended for multiple paths, Peng et al. \cite{peng2014multipath} pointed out their insufficiency. Thus, there is a need for designing new tools and frameworks for examining the performance of multipath congestion control algorithms. 
\end{itemize}

Congestion control was originally designed for  the transport layer. However, with the increasing scalability and robustness requirements, it has been extended to application layer (for real-time applications) and network layer (for controlling congestion at interim routers). As a result, \textcolor{black}{cross-layer} interactions have become critical and is the topic of our next discussion. 


\subsection{\textcolor{black}{Cross-Layer} Multipath Interactions}
\textcolor{black}{Cross-layer interactions between application layer-transport layer and transport layer-network layer opens up new horizons. The application layer and the transport layer can work together to design better protocols for real-time applications while the interaction between the transport layer and the network layer is simultaneously the most interesting and the most challenging problem.} Due to its importance, the rest of this subsection will focus only on transport layer-network layer interaction.

\textcolor{black}{Transport-layer} multipathing reduces packet reordering and is best suited for load balancing \cite{kelly2005stability}. This way, the transport layer can take some stress away from network layer, leaving network layer more scalable and robust \cite{arkko2009dagstuhl}. Network layer can also work in harmony with the transport layer to provide various benefits including minimizing congestion, differentiating between losses due to transmission or congestion in wireless environments etc. Due to this promise of numerous benefits, \textcolor{black}{cross-layer} interaction between the transport layer and the network layer has become a hot area of research. 
\textcolor{black}{Some examples of the cross-layer interaction between transport layer-network layer include MPTCP/OpenFlow \cite{van2013experiences} (wherein efficient network utilization is enabled due to the interaction between MPTCP and OpenFlow) and augmented MPTCP (A-MPTCP) \cite{coudron2013cross} (that is basically a \textcolor{black}{cross-layer} cooperation between MPTCP and location/ identification separation protocol \cite{farinacci2013locator}).}
The key challenges associated with \textcolor{black}{cross-layer} interaction of the transport layer and the network layer are described below.

\begin{itemize}
\item \textit{Heterogeneous convergence time.}
For designing \textcolor{black}{cross-layer} architectures, it is important to take into account the convergence time of both congestion control (depends on RTT, congestion window size) and routing (depends on operational costs, etc) algorithms. The stability and optimality of \textcolor{black}{cross-layer} architecture has been studied in literature by extending utility maximization framework\cite{he2006tcp, wang2005cross, anderson2003stability, he2007towards}. With increasing trend towards multipathing, this area is ripe for research.

\item \textit{Mitigating the tussle between various \textcolor{black}{stakeholders}.} There can be a difference in the design goals of network operators and end systems, this problem is commonly referred in literature as ``tussle" \cite{clark2002tussle}.  To truly reap the benefits of multipathing, \textcolor{black}{the} goals of network operators and end systems should be aligned. The \textcolor{black}{T}rilogy architecture \cite{abt2009trilogy} is aimed at minimizing this tussle and serves as the basis for future research in this area. 
\end{itemize}




\subsection{Multipathing and Green Networking}
While multipathing is envisioned to revolutionize wireless networks and datacenters, power consumption with the use of multiple paths is very high. Researchers are actively working towards making multipathing more energy efficient and environment friendly. Some prominent multipath-based green networking efforts that will pave the way for future research are discussed next.
 
\begin{itemize}
\item \textit{Wireless networks.}
Since mobile devices are power constrained so, a balanced trade-off is required between power consumption and use of multiple paths. Lim et al. \cite{lim2014improving, lim2014green} studied energy considerations for MPTCP on a mobile device (enabled with both cellular and Wi-Fi interfaces) and developed an energy efficient MPTCP protocol called eMPTCP. It was experimentally found that eMPTCP reduced power consumption by a significant percentage in comparison to MPTCP while providing sufficient  redundancy and fault tolerance. Energy efficiency of eMPTCP was further verified in \cite{limvalidation}. It is anticipated that eMPTCP will provide a firm basis for future efforts towards more energy efficient protocols. 

\item \textit{Datacenters.}
Large volumes of electricity is consumed by the datacenters and many efforts (\textcolor{black}{such as}  use of cooling towers, etc) have been made to reduce energy consumed by datacenters. Multipathing can also contribute towards the goal of green datacenters by enabling virtual machine (VM) migration. The seamless VM migration within and between datacenters provides the ability to benefit from energy efficiency \cite{nicutar2013evolving}.  
\end{itemize}

The research in the field of multipathing is in no sense limited to the above-mentioned four cases, there are few other areas that require sincere efforts \textcolor{black}{(e.g., how to effectively deploy diversity, development of flow scheduling mechanisms for reducing packet reordering at receiver)}. Researchers are keenly working towards overcoming these challenges to enable wide spread deployment of multipathing.


\section{Conclusions}
\label{Conclusion}
This paper explained in detail the transition from single-path to multipath support at the transport layer \textcolor{black}{by describing the modifications in traditional \textcolor{black}{connection setup}, flow control, sequence number splitting, ACK, and flow scheduling mechanisms.}
\textcolor{black}{Because rate control is the most attractive feature of transport layer, its mathematical model has also been discussed \textcolor{black}{with} an analysis on the stability and equilibrium properties.} Following this, the \textcolor{black}{reshaping} of congestion control algorithms for use with multiple paths accompanied by their specific \textcolor{black}{window-increase/window-decrease} functions is presented. Multipath congestion control algorithms are also analyzed for their performance in terms of two parameters namely: TCP fairness and responsiveness. Towards the end, the paper highlights some open research problems and challenges to give a direction for future work.


\bibliographystyle{IEEEtran}
\bibliography{multipathing}

\begin{thebibliography}{100}
\providecommand{\url}[1]{#1}
\csname url@samestyle\endcsname
\providecommand{\newblock}{\relax}
\providecommand{\bibinfo}[2]{#2}
\providecommand{\BIBentrySTDinterwordspacing}{\spaceskip=0pt\relax}
\providecommand{\BIBentryALTinterwordstretchfactor}{4}
\providecommand{\BIBentryALTinterwordspacing}{\spaceskip=\fontdimen2\font plus
\BIBentryALTinterwordstretchfactor\fontdimen3\font minus
  \fontdimen4\font\relax}
\providecommand{\BIBforeignlanguage}[2]{{%
\expandafter\ifx\csname l@#1\endcsname\relax
\typeout{** WARNING: IEEEtran.bst: No hyphenation pattern has been}%
\typeout{** loaded for the language `#1'. Using the pattern for}%
\typeout{** the default language instead.}%
\else
\language=\csname l@#1\endcsname
\fi
#2}}
\providecommand{\BIBdecl}{\relax}
\BIBdecl

\bibitem{ye2008improving}
T.~Ye, D.~Veitch, and J.~Bolot, ``Improving wireless security through network
  diversity,'' \emph{ACM SIGCOMM Computer Communication Review}, vol.~39,
  no.~1, pp. 34--44, 2008.

\bibitem{kim2005improving}
K.-H. Kim and K.~G. Shin, ``Improving {TCP} performance over wireless networks
  with collaborative multi-homed mobile hosts,'' in \emph{Proceedings of the
  3rd international conference on Mobile systems, applications, and
  services}.\hskip 1em plus 0.5em minus 0.4em\relax ACM, 2005, pp. 107--120.

\bibitem{fastestTCP}
C.~Paasch, G.~Detal, S.~Barre, F.~Duchene, and O.~Bonaventure, ``The fastest
  {TCP} connection with multipath {TCP},'' \emph{Available:
  \url{http://multipath-tcp.org/pmwiki.php?n=Main.50Gbps}}, 2013.

\bibitem{qadir2015exploiting}
J.~Qadir, A.~Ali, K.-L.~A. Yau, A.~Sathiaseelan, and J.~Crowcroft, ``Exploiting
  the power of multiplicity: a holistic survey of network-layer multipath,''
  \emph{IEEE Communications Surveys Tutorials}, 2015.

\bibitem{singhsurvey}
S.~Singh, T.~Das, and A.~Jukan, ``{A Survey on Internet Multipath Routing and
  Provisioning},'' \emph{IEEE Communications Surveys Tutorials}, 2015.

\bibitem{paasch2012exploring}
C.~Paasch, G.~Detal, F.~Duchene, C.~Raiciu, and O.~Bonaventure, ``Exploring
  mobile/{WiFi} handover with multipath {TCP},'' in \emph{Proceedings of the
  2012 ACM SIGCOMM workshop on Cellular networks: operations, challenges, and
  future design}.\hskip 1em plus 0.5em minus 0.4em\relax ACM, 2012, pp. 31--36.

\bibitem{iOS}
iOS: Multipath TCP Support~in iOS7,
  ``https://support.apple.com/en-us/ht201373.''

\bibitem{agache2012grin}
A.~Agache and C.~Raiciu, ``{GRIN}: utilizing the empty half of full bisection
  networks,'' in \emph{Proceedings of the 4th USENIX conference on Hot Topics
  in Cloud Ccomputing}.\hskip 1em plus 0.5em minus 0.4em\relax USENIX
  Association, 2012, pp. 7--7.

\bibitem{raiciu2011improving}
C.~Raiciu, S.~Barre, C.~Pluntke, A.~Greenhalgh, D.~Wischik, and M.~Handley,
  ``Improving datacenter performance and robustness with multipath {TCP},'' in
  \emph{ACM SIGCOMM Computer Communication Review}, vol.~41, no.~4.\hskip 1em
  plus 0.5em minus 0.4em\relax ACM, 2011, pp. 266--277.

\bibitem{abt2009trilogy}
M.~H. aBT Innovate, ``{The Trilogy Architecture for the Future Internet},''
  \emph{Towards the Future Internet: A European Research Perspective}, p.~79,
  2009.

\bibitem{kelly2005stability}
F.~Kelly and T.~Voice, ``Stability of end-to-end algorithms for joint routing
  and rate control,'' \emph{ACM SIGCOMM Computer Communication Review},
  vol.~35, no.~2, pp. 5--12, 2005.

\bibitem{dong2007multi}
Y.~Dong, D.~Wang, N.~Pissinou, and J.~Wang, ``Multi-path load balancing in
  transport layer,'' in \emph{Next Generation Internet Networks, 3rd EuroNGI
  Conference on}.\hskip 1em plus 0.5em minus 0.4em\relax IEEE, 2007, pp.
  135--142.

\bibitem{wischik2008resource}
D.~Wischik, M.~Handley, and M.~B. Braun, ``The resource pooling principle,''
  \emph{ACM SIGCOMM Computer Communication Review}, vol.~38, no.~5, pp. 47--52,
  2008.

\bibitem{savage1999end}
S.~Savage, A.~Collins, E.~Hoffman, J.~Snell, and T.~Anderson, ``{T}he
  end-to-end effects of {I}nternet path selection,'' in \emph{ACM SIGCOMM
  Computer Communication Review}, vol.~29, no.~4.\hskip 1em plus 0.5em minus
  0.4em\relax ACM, 1999, pp. 289--299.

\bibitem{apostolopoulos2000reliable}
J.~G. Apostolopoulos, ``Reliable video communication over lossy packet networks
  using multiple state encoding and path diversity,'' in \emph{Photonics West
  2001-Electronic Imaging}.\hskip 1em plus 0.5em minus 0.4em\relax
  International Society for Optics and Photonics, 2000, pp. 392--409.

\bibitem{liang2001real}
Y.~J. Liang, E.~G. Steinbach, and B.~Girod, ``Real-time voice communication
  over the internet using packet path diversity,'' in \emph{Proceedings of the
  ninth ACM international conference on Multimedia}.\hskip 1em plus 0.5em minus
  0.4em\relax ACM, 2001, pp. 431--440.

\bibitem{hossain20135g}
S.~Hossain, ``5g wireless communication systems,'' \emph{American Journal of
  Engineering Research (AJER) e-ISSN}, pp. 2320--0847, 2013.

\bibitem{zhuang2012multipath}
W.~Zhuang, N.~Mohammadizadeh, and X.~Shen, ``Multipath transmission for
  wireless {Internet} access---from an end-to-end transport layer
  perspective,'' \emph{J. Internet Technol}, vol.~13, no.~1, pp. 1--18, 2012.

\bibitem{ramaboli2012bandwidth}
A.~L. Ramaboli, O.~E. Falowo, and A.~H. Chan, ``Bandwidth aggregation in
  heterogeneous wireless networks: A survey of current approaches and issues,''
  \emph{Journal of Network and Computer Applications}, vol.~35, no.~6, pp.
  1674--1690, 2012.

\bibitem{habak2013bandwidth}
K.~Habak, K.~A. Harras, and M.~Youssef, ``Bandwidth aggregation techniques in
  heterogeneous multi-homed devices: A survey,'' \emph{arXiv preprint
  arXiv:1309.0542}, 2013.

\bibitem{addepalli2013heterogeneous}
S.~Addepalli, H.~G. Schulzrinne, A.~Singh, and G.~Ormazabal, ``Heterogeneous
  access: Survey and design considerations,'' 2013.

\bibitem{honda2011still}
M.~Honda, Y.~Nishida, C.~Raiciu, A.~Greenhalgh, M.~Handley, and H.~Tokuda, ``Is
  it still possible to extend {TCP}?'' in \emph{Proceedings of the 2011 ACM
  SIGCOMM conference on Internet measurement conference}.\hskip 1em plus 0.5em
  minus 0.4em\relax ACM, 2011, pp. 181--194.

\bibitem{sherry2012making}
J.~Sherry, S.~Hasan, C.~Scott, A.~Krishnamurthy, S.~Ratnasamy, and V.~Sekar,
  ``Making middleboxes someone else's problem: network processing as a cloud
  service,'' \emph{ACM SIGCOMM Computer Communication Review}, vol.~42, no.~4,
  pp. 13--24, 2012.

\bibitem{hesmans2013tcp}
B.~Hesmans, F.~Duchene, C.~Paasch, G.~Detal, and O.~Bonaventure, ``Are {TCP}
  extensions middlebox-proof?'' in \emph{Proceedings of the 2013 workshop on
  Hot topics in middleboxes and network function virtualization}.\hskip 1em
  plus 0.5em minus 0.4em\relax ACM, 2013, pp. 37--42.

\bibitem{postel1980rfc}
J.~Postel, ``{User Datagram Protocol} ({UDP}),'' \emph{IETF RFC 768}, 1980.

\bibitem{postel1981transmission}
------, ``Transmission control protocol,'' \emph{IETF RFC 793}, 1981.

\bibitem{jacobson2003rtp}
V.~Jacobson, R.~Frederick, S.~Casner, and H.~Schulzrinne, ``{RTP}: A transport
  protocol for real-time applications,'' \emph{IETF RFC 3550}, 2003.

\bibitem{peterson2007computer}
L.~L. Peterson and B.~S. Davie, \emph{Computer networks: a systems
  approach}.\hskip 1em plus 0.5em minus 0.4em\relax Elsevier, 2007.

\bibitem{floyd2000extension}
S.~Floyd, J.~Mahdavi, M.~Podolsky, and M.~Mathis, ``{A}n extension to the
  selective acknowledgement ({SACK}) option for {TCP},'' 2000.

\bibitem{braden1989requirements}
R.~Braden, ``Requirements for internet hosts-communication layers,'' 1989.

\bibitem{sripanidkulchai2004analysis}
K.~Sripanidkulchai, B.~Maggs, and H.~Zhang, ``An analysis of live streaming
  workloads on the internet,'' in \emph{Proceedings of the 4th ACM SIGCOMM
  conference on Internet measurement}.\hskip 1em plus 0.5em minus 0.4em\relax
  ACM, 2004, pp. 41--54.

\bibitem{van2002streaming}
J.~Van~der Merwe, S.~Sen, and C.~Kalmanek, ``Streaming video traffic:
  Characterization and network impact,'' in \emph{Proceedings of the Seventh
  International Web Content Caching and Distribution Workshop}, 2002.

\bibitem{wang2001empirical}
Y.~Wang, M.~Claypool, and Z.~Zuo, ``An empirical study of realvideo performance
  across the internet,'' in \emph{Proceedings of the 1st ACM SIGCOMM Workshop
  on Internet Measurement}.\hskip 1em plus 0.5em minus 0.4em\relax ACM, 2001,
  pp. 295--309.

\bibitem{jacobson1988congestion}
V.~Jacobson, ``Congestion avoidance and control,'' in \emph{ACM SIGCOMM
  computer communication review}, vol.~18, no.~4.\hskip 1em plus 0.5em minus
  0.4em\relax ACM, 1988, pp. 314--329.

\bibitem{padhye2000modeling}
J.~Padhye, V.~Firoiu, D.~F. Towsley, and J.~F. Kurose, ``Modeling {TCP} {Reno}
  performance: a simple model and its empirical validation,'' \emph{IEEE/ACM
  Transactions on Networking (ToN)}, vol.~8, no.~2, pp. 133--145, 2000.

\bibitem{brakmo1995tcp}
L.~S. Brakmo and L.~L. Peterson, ``{TCP} {V}egas: {End} to end congestion
  avoidance on a global {Internet},'' \emph{Selected Areas in Communications,
  IEEE Journal on}, vol.~13, no.~8, pp. 1465--1480, 1995.

\bibitem{liu2008tcp}
S.~Liu, T.~Ba{\c{s}}ar, and R.~Srikant, ``{TCP}-{I}llinois: A loss-and
  delay-based congestion control algorithm for high-speed networks,''
  \emph{Performance Evaluation}, vol.~65, no.~6, pp. 417--440, 2008.

\bibitem{gevros2001congestion}
P.~Gevros, J.~Crowcroft, P.~Kirstein, and S.~Bhatti, ``Congestion control
  mechanisms and the best effort service model,'' \emph{Network, IEEE},
  vol.~15, no.~3, pp. 16--26, 2001.

\bibitem{floyd1993random}
S.~Floyd and V.~Jacobson, ``Random early detection gateways for congestion
  avoidance,'' \emph{Networking, IEEE/ACM Transactions on}, vol.~1, no.~4, pp.
  397--413, 1993.

\bibitem{athuraliya2000random}
S.~Athuraliya, S.~Low, and D.~Lapsley, ``Random early marking,'' in
  \emph{Quality of Future Internet Services}.\hskip 1em plus 0.5em minus
  0.4em\relax Springer, 2000, pp. 43--54.

\bibitem{kunniyur2001analysis}
S.~Kunniyur and R.~Srikant, ``Analysis and design of an adaptive virtual queue
  ({AVQ}) algorithm for active queue management,'' in \emph{ACM SIGCOMM
  Computer Communication Review}, vol.~31, no.~4.\hskip 1em plus 0.5em minus
  0.4em\relax ACM, 2001, pp. 123--134.

\bibitem{nichols2012controlling}
K.~Nichols and V.~Jacobson, ``Controlling queue delay,'' \emph{Communications
  of the ACM}, vol.~55, no.~7, pp. 42--50, 2012.

\bibitem{kelly1998rate}
F.~P. Kelly, A.~K. Maulloo, and D.~K. Tan, ``Rate control for communication
  networks: shadow prices, proportional fairness and stability,'' \emph{Journal
  of the Operational Research society}, vol.~49, no.~3, pp. 237--252, 1998.

\bibitem{nash1950bargaining}
J.~F. Nash~Jr, ``The bargaining problem,'' \emph{Econometrica: Journal of the
  Econometric Society}, pp. 155--162, 1950.

\bibitem{chiang2010fairness}
T.~Lan, D.~Kao, M.~Chiang, and A.~Sabharwal, ``An axiomatic theory of fairness
  in network resource allocation,'' in \emph{INFOCOM, 2010 Proceedings IEEE},
  March 2010, pp. 1--9.

\bibitem{kelly1997charging}
F.~Kelly, ``Charging and rate control for elastic traffic,'' \emph{European
  transactions on Telecommunications}, vol.~8, no.~1, pp. 33--37, 1997.

\bibitem{thomas1980decentralized}
{Thomas J. Watson IBM Research Center. Research Division and Jaffe, JM},
  \emph{A Decentralized,`` optimal,'' Multiple-user Flow Control Algorithm},
  1980.

\bibitem{floyd1999promoting}
S.~Floyd and K.~Fall, ``Promoting the use of end-to-end congestion control in
  the internet,'' \emph{IEEE/ACM Transactions on Networking (TON)}, vol.~7,
  no.~4, pp. 458--472, 1999.

\bibitem{mo2000fair}
J.~Mo and J.~Walrand, ``Fair end-to-end window-based congestion control,''
  \emph{IEEE/ACM Transactions on Networking (ToN)}, vol.~8, no.~5, pp.
  556--567, 2000.

\bibitem{kelly2000models}
F.~P. Kelly, ``Models for a self--managed {Internet},'' \emph{Philosophical
  Transactions of the Royal Society of London A: Mathematical, Physical and
  Engineering Sciences}, vol. 358, no. 1773, pp. 2335--2348, 2000.

\bibitem{kelly2001mathematical}
{F. P. Kelly}, ``Mathematical modelling of the {Internet},'' \emph{Mathematics
  unlimited-2001 and beyond}, pp. 685--702, 2001.

\bibitem{shakkottai2008network}
S.~Shakkottai, S.~G. Shakkottai, and R.~Srikant, \emph{Network optimization and
  control}.\hskip 1em plus 0.5em minus 0.4em\relax Now Publishers Inc, 2008.

\bibitem{kelly2009resource}
F.~P. Kelly, L.~Massouli{\'e}, and N.~S. Walton, ``Resource pooling in
  congested networks: proportional fairness and product form,'' \emph{Queueing
  Systems}, vol.~63, no. 1-4, pp. 165--194, 2009.

\bibitem{srikant2004mathematics}
R.~Srikant, \emph{The mathematics of Internet congestion control}.\hskip 1em
  plus 0.5em minus 0.4em\relax Springer, 2004.

\bibitem{eun2007limitation}
D.~Y. Eun, ``On the limitation of fluid-based approach for {Internet}
  congestion control,'' \emph{Telecommunication Systems}, vol.~34, pp. 3--11,
  2007.

\bibitem{low2002internet}
S.~H. Low, F.~Paganini, and J.~C. Doyle, ``Internet congestion control,''
  \emph{Control Systems, IEEE}, vol.~22, no.~1, pp. 28--43, 2002.

\bibitem{key2011path}
P.~Key, L.~Massouli{\'e}, and D.~Towsley, ``Path selection and multipath
  congestion control,'' \emph{communications of the acm}, vol.~54, no.~1, pp.
  109--116, 2011.

\bibitem{stewart2000stream}
R.~Stewart, Q.~Xie, K.~Morneault, C.~Sharp, H.~Schwarzbauer, T.~Taylor,
  I.~Rytina, M.~Kalla, L.~Zhang, and V.~Paxson, ``{Stream Control Transmission
  Protocol},'' \emph{IETF RFC 2960}, 2000.

\bibitem{stewart2007stream}
R.~Stewart, ``Stream control transmission protocol,'' \emph{IETF RFC 4960},
  2007.

\bibitem{stewart2002sctp}
R.~Stewart, M.~Ramalho, Q.~Xie, M.~Tuexen, and P.~Conrad, ``{SCTP} partial
  reliability extension,'' \emph{IETF RFC 3758}, 2002.

\bibitem{stewart2004rfc}
{Stewart, R and Ramalho, M and Xie, Q and Tuexen, M and Conrad, P}, ``{RFC
  3758: Stream control transmission protocol (SCTP) partial reliability
  extension},'' \emph{Request for Comments, IETF}, vol.~8, 2004.

\bibitem{argyriou2003bandwidth}
A.~Argyriou and V.~Madisetti, ``Bandwidth aggregation with {SCTP},'' in
  \emph{Global Telecommunications Conference, 2003. GLOBECOM'03. IEEE},
  vol.~7.\hskip 1em plus 0.5em minus 0.4em\relax IEEE, 2003, pp. 3716--3721.

\bibitem{stewart2003stream}
R.~Stewart, M.~Ramalho, Q.~Xie, M.~Tuexen, I.~Rytina, M.~Belinchon, and
  P.~Conrad, ``{Stream Control Transmission Protocol {(SCTP)} Dynamic Address
  Reconfiguration}" draft-ietf-tsvwg-addip-sctp-07. txt,'' \emph{IETF,
  September}, 2003.

\bibitem{stewart2005stream}
R.~Stewart, M.~Ramalho, Q.~Xie, M.~Tuexen, and P.~Conrad, ``{Stream Control
  Transmission Protocol (SCTP) Dynamic Address Reconfiguration}.
  draft-ietf-tsvwgaddip-sctp-11,'' txt, Internet Draft (work in progress),
  IETF, Tech. Rep., 2005.

\bibitem{maruyama2007stream}
S.~Maruyama, M.~Tuexen, R.~Stewart, Q.~Xie, and M.~Kozuka, ``{Stream Control
  Transmission Protocol (SCTP) Dynamic Address Reconfiguration},'' 2007.

\bibitem{casetti2004westwood}
C.~Casetti and W.~Gaiotto, ``Westwood {SCTP}: load balancing over multipaths
  using bandwidth-aware source scheduling,'' in \emph{Vehicular Technology
  Conference, 2004. VTC2004-Fall. 2004 IEEE 60th}, vol.~4.\hskip 1em plus 0.5em
  minus 0.4em\relax IEEE, 2004, pp. 3025--3029.

\bibitem{abd2004ls}
A.~Abd El~Al, T.~Saadawi, and M.~Lee, ``{LS-SCTP}: a bandwidth aggregation
  technique for stream control transmission protocol,'' \emph{Computer
  Communications}, vol.~27, no.~10, pp. 1012--1024, 2004.

\bibitem{abd2004improving}
A.~Abd, T.~Saadawi, M.~Lee \emph{et~al.}, ``Improving throughput and
  reliability in mobile wireless networks via transport layer bandwidth
  aggregation,'' \emph{Computer Networks}, vol.~46, no.~5, pp. 635--649, 2004.

\bibitem{becke2010load}
M.~Becke, T.~Dreibholz, J.~Iyengar, P.~Natarajan, and M.~Tuexen, ``Load sharing
  for the stream control transmission protocol ({SCTP}),''
  \emph{draft-tuexen-tsvwg-sctpmultipath-01. txt, Dec}, 2010.

\bibitem{amer2013load}
P.~Amer, M.~Becke, T.~Dreibholz, N.~Ekiz, J.~Iyengar, P.~Natarajan, R.~Stewart,
  and M.~T{\"u}xen, ``Load sharing for the stream control transmission protocol
  {(SCTP)},'' \emph{IETF ID: draft-tuexen-tsvwg-sctp-multipath-06 (work in
  progress)}, 2013.

\bibitem{koh2005mobile}
S.~J. Koh, Q.~Xie, and S.~D. Park, ``Mobile sctp (msctp) for ip handover
  support,'' \emph{IETF Draft, draft-sjkoh-msctp-01 (October 2005)}, 2005.

\bibitem{riegel2007mobile}
M.~Riegel \emph{et~al.}, ``Mobile sctp. draft-riegel-tuexen-mobile-sctp-09,''
  \emph{URL: http://www. watersprings.
  org/pub/id/draft-riegeltuexen-mobile-sctp-05. txt (19 November 2009)}, 2007.

\bibitem{iyengar2006concurrent}
J.~R. Iyengar, P.~D. Amer, and R.~Stewart, ``Concurrent multipath transfer
  using ({SCTP}) multihoming over independent end-to-end paths,''
  \emph{Networking, IEEE/ACM Transactions on}, vol.~14, no.~5, pp. 951--964,
  2006.

\bibitem{huang2007wimp}
C.-M. Huang and C.-H. Tsai, ``{W}i{MP}-{SCTP}: Multi-path transmission using
  stream control transmission protocol ({SCTP}) in wireless networks,'' in
  \emph{Advanced Information Networking and Applications Workshops, 2007,
  AINAW'07. 21st International Conference on}, vol.~1.\hskip 1em plus 0.5em
  minus 0.4em\relax IEEE, 2007, pp. 209--214.

\bibitem{liao2008cmpsctp}
J.~Liao, J.~Wang, and X.~Zhu, ``{cmpSCTP}: An extension of {SCTP} to support
  concurrent multi-path transfer,'' in \emph{Communications, 2008. ICC'08. IEEE
  International Conference on}.\hskip 1em plus 0.5em minus 0.4em\relax IEEE,
  2008, pp. 5762--5766.

\bibitem{budzisz2009concurrent}
{\L}.~Budzisz, R.~Ferr{\'u}s, F.~Casadevall, and P.~Amer, ``On concurrent
  multipath transfer in {SCTP}-based handover scenarios,'' in
  \emph{Communications, 2009. ICC'09. IEEE International Conference on}.\hskip
  1em plus 0.5em minus 0.4em\relax IEEE, 2009, pp. 1--6.

\bibitem{mirani2010data}
F.~H. Mirani, N.~Boukhatem, and M.~A. Tran, ``A data-scheduling mechanism for
  multi-homed mobile terminals with disparate link latencies,'' in
  \emph{Vehicular Technology Conference Fall (VTC 2010-Fall), 2010 IEEE
  72nd}.\hskip 1em plus 0.5em minus 0.4em\relax IEEE, 2010, pp. 1--5.

\bibitem{yuan2010extension}
Y.~Yuan, Z.~Zhang, J.~Li, J.~Shi, J.~Zhou, G.~Fang, and E.~Dutkiewicz,
  ``Extension of {SCTP} for concurrent multi-path transfer with parallel
  subflows,'' in \emph{Wireless Communications and Networking Conference
  (WCNC), 2010 IEEE}.\hskip 1em plus 0.5em minus 0.4em\relax IEEE, 2010, pp.
  1--6.

\bibitem{ford2011architectural}
A.~Ford, C.~Raiciu, M.~Handley, S.~Barre, J.~Iyengar \emph{et~al.},
  ``Architectural guidelines for multipath {TCP} development,'' \emph{IETF,
  Informational RFC}, vol. 6182, pp. 2070--1721, 2011.

\bibitem{raiciu2012hard}
C.~Raiciu, C.~Paasch, S.~Barre, A.~Ford, M.~Honda, F.~Duchene, O.~Bonaventure,
  M.~Handley \emph{et~al.}, ``How hard can it be? designing and implementing a
  deployable multipath {TCP}.'' in \emph{NSDI}, vol.~12, 2012, pp. 29--29.

\bibitem{ford2013tcp}
A.~Ford, C.~Raiciu, M.~Handley, and O.~Bonaventure, ``{TCP} extensions for
  multipath operation with multiple addresses,'' Tech. Rep., 2013.

\bibitem{li2012network}
M.~Li, A.~Lukyanenko, and Y.~Cui, ``Network coding based multipath {TCP},'' in
  \emph{Computer Communications Workshops (INFOCOM WKSHPS), 2012 IEEE
  Conference on}.\hskip 1em plus 0.5em minus 0.4em\relax IEEE, 2012, pp.
  25--30.

\bibitem{diop2012qos}
C.~Diop, G.~Dugu{\'e}, C.~Chassot, and E.~Exposito, ``{QoS}-oriented {MPTCP}
  extensions for multimedia multi-homed systems,'' in \emph{Advanced
  Information Networking and Applications Workshops (WAINA), 2012 26th
  International Conference on}.\hskip 1em plus 0.5em minus 0.4em\relax IEEE,
  2012, pp. 1119--1124.

\bibitem{li2013delayed}
M.~Li, A.~Lukyanenko, S.~Tarkoma, and A.~Yla-Jaaski, ``{T}he {D}elayed {ACK}
  evolution in {MPTCP},'' in \emph{Global Communications Conference (GLOBECOM),
  2013 IEEE}.\hskip 1em plus 0.5em minus 0.4em\relax IEEE, 2013, pp.
  2282--2288.

\bibitem{van2013experiences}
R.~van~der Pol, M.~Bredel, A.~Barczyk, B.~Overeinder, N.~van Adrichem, and
  F.~Kuipers, ``Experiences with {MPTCP} in an intercontinental openflow
  network,'' in \emph{Proceedings of the 29th TERENA Network Conference
  (TNC2013)}, 2013.

\bibitem{coudron2013cross}
M.~Coudron, S.~Secci, G.~Pujolle, P.~Raad, and P.~Gallard, ``Cross-layer
  cooperation to boost multipath {TCP} performance in cloud networks,'' in
  \emph{Cloud Networking (CloudNet), 2013 IEEE 2nd International Conference
  on}.\hskip 1em plus 0.5em minus 0.4em\relax IEEE, 2013, pp. 58--66.

\bibitem{zhou2013goodput}
D.~Zhou, W.~Song, and M.~Shi, ``Goodput improvement for multipath {TCP} by
  congestion window adaptation in multi-radio devices,'' in \emph{Consumer
  Communications and Networking Conference (CCNC), 2013 IEEE}.\hskip 1em plus
  0.5em minus 0.4em\relax IEEE, 2013, pp. 508--514.

\bibitem{li2013tolerating}
M.~Li, A.~Lukyanenko, S.~Tarkoma, Y.~Cui, and A.~Yla-Jaaski, A.ki, ``Tolerating
  path heterogeneity in multipath {TCP} with bounded receive buffers,'' in
  \emph{ACM SIGMETRICS Performance Evaluation Review}, vol.~41, no.~1.\hskip
  1em plus 0.5em minus 0.4em\relax ACM, 2013, pp. 375--376.

\bibitem{yang2014using}
F.~Yang and P.~Amer, ``Using one-way communication delay for in-order arrival
  mptcp scheduling,'' \emph{Proceedings of the 9th International Conference on
  Communications and Networking in China (CHINACOM)}, pp. 122--125, 2014.

\bibitem{cui2015fmtcp}
Y.~Cui, L.~Wang, X.~Wang, H.~Wang, and Y.~Wang, ``{FMTCP}: A fountain
  code-based multipath transmission control protocol,'' \emph{Networking,
  IEEE/ACM Transactions on}, vol.~23, no.~2, pp. 465--478, 2015.

\bibitem{le2015forward}
T.-A. Le and L.~X. Bui, ``{Forward Delay-based Packet Scheduling Algorithm for
  Multipath TCP},'' \emph{arXiv preprint arXiv:1501.03196}, 2015.

\bibitem{magalhaes2001transport}
L.~Magalhaes and R.~Kravets, ``Transport level mechanisms for bandwidth
  aggregation on mobile hosts,'' in \emph{Network Protocols, 2001. Ninth
  International Conference on}.\hskip 1em plus 0.5em minus 0.4em\relax IEEE,
  2001, pp. 165--171.

\bibitem{lee2002improving}
Y.~Lee, I.~Park, and Y.~Choi, ``Improving {TCP} performance in multipath packet
  forwarding networks,'' \emph{Communications and Networks, Journal of},
  vol.~4, no.~2, pp. 148--157, 2002.

\bibitem{hsieh2002ptcp}
H.-Y. Hsieh and R.~Sivakumar, ``{pTCP: An end-to-end transport layer protocol
  for striped connections},'' in \emph{Network Protocols, 2002. Proceedings.
  10th IEEE International Conference on}.\hskip 1em plus 0.5em minus
  0.4em\relax IEEE, 2002, pp. 24--33.

\bibitem{hsieh2005transport}
{Hsieh, Hung-Yun and Sivakumar, Raghupathy}, ``A transport layer approach for
  achieving aggregate bandwidths on multi-homed mobile hosts,'' \emph{Wireless
  Networks}, vol.~11, no. 1-2, pp. 99--114, 2005.

\bibitem{hsieh2003receiver}
H.-Y. Hsieh, K.-H. Kim, Y.~Zhu, and R.~Sivakumar, ``A receiver-centric
  transport protocol for mobile hosts with heterogeneous wireless interfaces,''
  in \emph{Proceedings of the 9th annual international conference on Mobile
  computing and networking}.\hskip 1em plus 0.5em minus 0.4em\relax ACM, 2003,
  pp. 1--15.

\bibitem{kim2005receiver}
K.-H. Kim, Y.~Zhu, R.~Sivakumar, and H.-Y. Hsieh, ``A receiver-centric
  transport protocol for mobile hosts with heterogeneous wireless interfaces,''
  \emph{Wireless Networks}, vol.~11, no.~4, pp. 363--382, 2005.

\bibitem{cetinkaya2004opportunistic}
C.~Cetinkaya and E.~W. Knightly, ``Opportunistic traffic scheduling over
  multiple network paths,'' in \emph{INFOCOM 2004. Twenty-third AnnualJoint
  Conference of the IEEE Computer and Communications Societies}, vol.~3.\hskip
  1em plus 0.5em minus 0.4em\relax IEEE, 2004, pp. 1928--1937.

\bibitem{zhang2004transport}
M.~Zhang, J.~Lai, A.~Krishnamurthy, L.~L. Peterson, and R.~Y. Wang, ``A
  transport layer approach for improving end-to-end performance and robustness
  using redundant paths.'' in \emph{USENIX Annual Technical Conference, General
  Track}, 2004, pp. 99--112.

\bibitem{chen2004multipath}
J.~Chen, K.~Xu, and M.~Gerla, ``{Multipath TCP in lossy wireless
  environment},'' in \emph{Proc. IFIP Third Annual Mediterranean Ad Hoc
  Networking Workshop (Med-Hoc-Net'04)}, 2004, pp. 263--270.

\bibitem{rojviboonchai2004evaluation}
K.~Rojviboonchai and A.~Hitoshi, ``An evaluation of multi-path transmission
  control protocol ({M}/{TCP}) with robust acknowledgement schemes,''
  \emph{IEICE transactions on communications}, vol.~87, no.~9, pp. 2699--2707,
  2004.

\bibitem{rojviboonchai2005rm}
K.~Rojviboonchai, T.~Osuga, and H.~Aida, ``{RM}/{TCP}: Protocol for reliable
  multi-path transport over the internet,'' in \emph{Advanced Information
  Networking and Applications, 2005. AINA 2005. 19th International Conference
  on}, vol.~1.\hskip 1em plus 0.5em minus 0.4em\relax IEEE, 2005, pp. 801--806.

\bibitem{sarkar2006concurrent}
D.~Sarkar, ``A concurrent multipath {TCP} and its markov model,'' in
  \emph{Communications, 2006. ICC'06. IEEE International Conference on},
  vol.~2.\hskip 1em plus 0.5em minus 0.4em\relax IEEE, 2006, pp. 615--620.

\bibitem{sarkar2006qrp04}
D.~Sarkar and S.~Paul, ``{QRP04-3}: Architecture, implementation, and
  evaluation of cmptcp westwood,'' in \emph{Global Telecommunications
  Conference, 2006. GLOBECOM'06. IEEE}.\hskip 1em plus 0.5em minus 0.4em\relax
  IEEE, 2006, pp. 1--5.

\bibitem{dong2007concurrency}
Y.~Dong, N.~Pissinou, and J.~Wang, ``Concurrency handling in {TCP},'' in
  \emph{Communication Networks and Services Research, 2007. CNSR'07. Fifth
  Annual Conference on}.\hskip 1em plus 0.5em minus 0.4em\relax IEEE, 2007, pp.
  255--262.

\bibitem{sharma2008mplot}
V.~Sharma, S.~Kalyanaraman, K.~Kar, K.~Ramakrishnan, and V.~Subramanian,
  ``{MPLOT}: A transport protocol exploiting multipath diversity using erasure
  codes,'' in \emph{INFOCOM 2008. The 27th Conference on Computer
  Communications. IEEE}.\hskip 1em plus 0.5em minus 0.4em\relax IEEE, 2008.

\bibitem{sharma2012transport}
V.~Sharma, K.~Kar, K.~Ramakrishnan, and S.~Kalyanaraman, ``A transport protocol
  to exploit multipath diversity in wireless networks,'' \emph{IEEE/ACM
  Transactions on Networking (TON)}, vol.~20, no.~4, pp. 1024--1039, 2012.

\bibitem{budzisz2008towards}
{\L}.~Budzisz, R.~Ferr{\'u}s, A.~Brunstrom, K.-J. Grinnemo, R.~Fracchia,
  G.~Galante, and F.~Casadevall, ``Towards transport-layer mobility: Evolution
  of {SCTP} multihoming,'' \emph{Computer Communications}, vol.~31, no.~5, pp.
  980--998, 2008.

\bibitem{jungmaier2006sctp}
A.~Jungmaier and E.~P. Rathgeb, ``On {SCTP} multi-homing performance,''
  \emph{Telecommunication Systems}, vol.~31, no. 2-3, pp. 141--161, 2006.

\bibitem{Iyengar2006}
J.~R. Iyengar, P.~D. Amer, and R.~Stewart, ``Concurrent multipath transfer
  using ({SCTP}) multihoming over independent end-to-end paths,''
  \emph{Networking, IEEE/ACM Transactions on}, vol.~14, no.~5, pp. 951--964,
  2006.

\bibitem{fu2004sctp}
S.~Fu and M.~Atiquzzaman, ``{SCTP}: State of the art in research, products, and
  technical challenges,'' \emph{Communications Magazine, IEEE}, vol.~42, no.~4,
  pp. 64--76, 2004.

\bibitem{budzisz2012taxonomy}
{\L}.~Budzisz, J.~Garcia, A.~Brunstrom, and R.~Ferr{\'u}s, ``A taxonomy and
  survey of sctp research,'' \emph{ACM Computing Surveys (CSUR)}, vol.~44,
  no.~4, p.~18, 2012.

\bibitem{paasch2014multipath}
C.~Paasch and O.~Bonaventure, ``Multipath {TCP},'' \emph{Communications of the
  ACM}, vol.~57, no.~4, pp. 51--57, 2014.

\bibitem{mao2006mrtp}
S.~Mao, D.~Bushmitch, S.~Narayanan, and S.~S. Panwar, ``{MRTP}: a multiflow
  real-time transport protocol for ad hoc networks,'' \emph{Multimedia, IEEE
  Transactions on}, vol.~8, no.~2, pp. 356--369, 2006.

\bibitem{singh2013mprtp}
V.~Singh, S.~Ahsan, and J.~Ott, ``{MPRTP}: multipath considerations for
  real-time media,'' in \emph{Proceedings of the 4th ACM Multimedia Systems
  Conference}.\hskip 1em plus 0.5em minus 0.4em\relax ACM, 2013, pp. 190--201.

\bibitem{zhang2014general}
W.~Zhang, W.~Lei, S.~Liu, and G.~Li, ``A general framework of multipath
  transport system based on application-level relay,'' \emph{Computer
  Communications}, 2014.

\bibitem{mitzenmacher2001power}
M.~Mitzenmacher, ``The power of two choices in randomized load balancing,''
  \emph{Parallel and Distributed Systems, IEEE Transactions on}, vol.~12,
  no.~10, pp. 1094--1104, 2001.

\bibitem{tullimas2008multimedia}
S.~Tullimas, T.~Nguyen, R.~Edgecomb, and S.-c. Cheung, ``Multimedia streaming
  using multiple {TCP} connections,'' \emph{ACM Transactions on Multimedia
  Computing, Communications, and Applications (TOMM)}, vol.~4, no.~2, p.~12,
  2008.

\bibitem{wang2009multipath}
B.~Wang, W.~Wei, Z.~Guo, and D.~Towsley, ``Multipath live streaming via {TCP}:
  scheme, performance and benefits,'' \emph{ACM Transactions on Multimedia
  Computing, Communications, and Applications (TOMCCAP)}, vol.~5, no.~3, p.~25,
  2009.

\bibitem{han2004overlay}
H.~Han, S.~Shakkottai, C.~Hollot, R.~Srikant, and D.~Towsley, ``Overlay {TCP}
  for multi-path routing and congestion control,'' in \emph{IMA Workshop on
  Measurements and Modeling of the Internet}, 2004.

\bibitem{peng2014multipath}
Q.~Peng, A.~Walid, J.-S. Hwang, and S.~H. Low, ``{Multipath TCP: Analysis,
  Design, and Implementation},'' 2014.

\bibitem{dreibholz2011impact}
T.~Dreibholz, M.~Becke, H.~Adhari, and E.~P. Rathgeb, ``{O}n the impact of
  congestion control for {C}oncurrent {M}ultipath {T}ransfer on the transport
  layer,'' in \emph{Telecommunications (ConTEL), Proceedings of the 2011 11th
  International Conference on}.\hskip 1em plus 0.5em minus 0.4em\relax IEEE,
  2011, pp. 397--404.

\bibitem{dreibholz2012simulation}
T.~Dreibholz, H.~Adhari, M.~Becke, and E.~P. Rathgeb, ``Simulation and
  experimental evaluation of multipath congestion control strategies,'' in
  \emph{Advanced Information Networking and Applications Workshops (WAINA),
  2012 26th International Conference on}.\hskip 1em plus 0.5em minus
  0.4em\relax IEEE, 2012, pp. 1113--1118.

\bibitem{khalili2013mptcp}
R.~Khalili, N.~Gast, M.~Popovic, and J.-Y. Le~Boudec, ``{MPTCP} is not
  pareto-optimal: performance issues and a possible solution,'' \emph{IEEE/ACM
  Transactions on Networking (TON)}, vol.~21, no.~5, pp. 1651--1665, 2013.

\bibitem{honda2009multipath}
M.~Honda, Y.~Nishida, L.~Eggert, P.~Sarolahti, and H.~Tokuda, ``Multipath
  congestion control for shared bottleneck,'' in \emph{Proc. PFLDNeT workshop},
  2009, pp. 19--24.

\bibitem{han2006multi}
H.~Han, S.~Shakkottai, C.~Hollot, R.~Srikant, and D.~Towsley, ``Multi-path
  {TCP}: a joint congestion control and routing scheme to exploit path
  diversity in the internet,'' \emph{IEEE/ACM Transactions on Networking
  (TON)}, vol.~14, no.~6, pp. 1260--1271, 2006.

\bibitem{wischik2011design}
D.~Wischik, C.~Raiciu, A.~Greenhalgh, and M.~Handley, ``Design, implementation
  and evaluation of congestion control for multipath {TCP},'' in
  \emph{Proceedings of the 8th USENIX conference on Networked systems design
  and implementation}.\hskip 1em plus 0.5em minus 0.4em\relax USENIX
  Association, 2011, pp. 8--8.

\bibitem{raiciu2011coupled}
C.~Raiciu, M.~Handley, and D.~Wischik, ``Coupled congestion control for
  multipath transport protocols,'' Tech. Rep., 2011.

\bibitem{hassayoun2011dynamic}
S.~Hassayoun, J.~Iyengar, and D.~Ros, ``Dynamic window coupling for multipath
  congestion control,'' in \emph{Network Protocols (ICNP), 2011 19th IEEE
  International Conference on}.\hskip 1em plus 0.5em minus 0.4em\relax IEEE,
  2011, pp. 341--352.

\bibitem{khalili2012non}
R.~Khalili, N.~G. Gast, M.~Popovic, U.~Upadhyay, and J.-Y. Le~Boudec,
  ``Non-pareto optimality of {MPTCP}: Performance issues and a possible
  solution,'' Tech. Rep., 2012.

\bibitem{singh2013enhancing}
A.~Singh, M.~Xiang, A.~Konsgen, C.~Goerg, and Y.~Zaki, ``Enhancing fairness and
  congestion control in multipath {TCP},'' in \emph{Wireless and Mobile
  Networking Conference (WMNC), 2013 6th Joint IFIP}.\hskip 1em plus 0.5em
  minus 0.4em\relax IEEE, 2013, pp. 1--8.

\bibitem{PengWL13}
Q.~Peng, A.~Walid, and S.~H. Low, ``{Multipath TCP: Analysis and Design},''
  \emph{CoRR}, vol. abs/1308.3119, 2013.

\bibitem{ming2014mptcp}
L.~Ming, A.~Lukyanenko, S.~Tarkoma, and A.~Yla-Jaaski, ``{MPTCP} incast in data
  center networks,'' \emph{Communications, China}, vol.~11, no.~4, pp. 25--37,
  2014.

\bibitem{hu2011research}
M.~Hu, J.-n. Gan, and Y.-n. Guo, ``The research of adaptive weighted congestion
  control in {MPTCP},'' in \emph{Electrical and Control Engineering (ICECE),
  2011 International Conference on}.\hskip 1em plus 0.5em minus 0.4em\relax
  IEEE, 2011, pp. 1350--1353.

\bibitem{wischik2010balancing}
D.~Wischik, C.~Raiciu, and M.~Handley, ``Balancing resource pooling and
  equipoise in multipath transport,'' \emph{Submitted to ACM SIGCOMM}, 2010.

\bibitem{scharfmultipath}
M.~Scharf, ``Multipath transport challenges and solutions.''

\bibitem{kostopoulos2010towards}
A.~Kostopoulos, H.~Warma, T.~Leva, B.~Heinrich, A.~Ford, and L.~Eggert,
  ``Towards multipath {TCP} adoption: challenges and opportunities,'' in
  \emph{Next Generation Internet (NGI), 2010 6th EURO-NF Conference on}.\hskip
  1em plus 0.5em minus 0.4em\relax IEEE, 2010, pp. 1--8.

\bibitem{saltzer1984end}
J.~H. Saltzer, D.~P. Reed, and D.~D. Clark, ``End-to-end arguments in system
  design,'' \emph{ACM Transactions on Computer Systems (TOCS)}, vol.~2, no.~4,
  pp. 277--288, 1984.

\bibitem{raiciu2013recent}
C.~Raiciu, J.~Iyengar, O.~Bonaventure \emph{et~al.}, ``Recent advances in
  reliable transport protocols,'' \emph{SIGCOMM ebook on Recent Advances in
  Networking}, 2013.

\bibitem{detal2013multipath}
G.~Detal, C.~Paasch, and O.~Bonaventure, ``{M}ultipath in the middle (box),''
  in \emph{Proceedings of the 2013 workshop on Hot topics in middleboxes and
  network function virtualization}.\hskip 1em plus 0.5em minus 0.4em\relax ACM,
  2013, pp. 1--6.

\bibitem{papadimitriou2011open}
D.~Papadimitriou, M.~Welzl, M.~Scharf, and B.~Briscoe, ``Open research issues
  in internet congestion control,'' Tech. Rep., 2011.

\bibitem{arkko2009dagstuhl}
J.~Arkko, B.~Briscoe, L.~Eggert, A.~Feldmann, and M.~Handley, ``Dagstuhl
  perspectives workshop on end-to-end protocols for the future internet,''
  \emph{ACM SIGCOMM Computer Communication Review}, vol.~39, no.~2, pp. 42--47,
  2009.

\bibitem{farinacci2013locator}
D.~Farinacci, D.~Lewis, D.~Meyer, and V.~Fuller, ``The locator/id separation
  protocol {(LISP)},'' 2013.

\bibitem{he2006tcp}
J.~He, M.~Chiang, and J.~Rexford, ``{TCP/IP interaction based on congestion
  price: Stability and optimality},'' in \emph{Communications, 2006. ICC'06.
  IEEE International Conference on}, vol.~3.\hskip 1em plus 0.5em minus
  0.4em\relax IEEE, 2006, pp. 1032--1039.

\bibitem{wang2005cross}
J.~Wang, L.~Li, S.~H. Low, and J.~C. Doyle, ``Cross-layer optimization in
  {TCP/IP} networks,'' \emph{Networking, IEEE/ACM Transactions on}, vol.~13,
  no.~3, pp. 582--595, 2005.

\bibitem{anderson2003stability}
E.~J. Anderson and T.~E. Anderson, ``On the stability of adaptive routing in
  the presence of congestion control,'' in \emph{INFOCOM 2003. Twenty-Second
  Annual Joint Conference of the IEEE Computer and Communications. IEEE
  Societies}, vol.~2.\hskip 1em plus 0.5em minus 0.4em\relax IEEE, 2003, pp.
  948--958.

\bibitem{he2007towards}
J.~He, M.~Bresler, M.~Chiang, and J.~Rexford, ``Towards robust multi-layer
  traffic engineering: Optimization of congestion control and routing,''
  \emph{Selected Areas in Communications, IEEE Journal on}, vol.~25, no.~5, pp.
  868--880, 2007.

\bibitem{clark2002tussle}
D.~D. Clark, J.~Wroclawski, K.~R. Sollins, and R.~Braden, ``Tussle in
  cyberspace: defining tomorrow's internet,'' in \emph{ACM SIGCOMM Computer
  Communication Review}, vol.~32, no.~4.\hskip 1em plus 0.5em minus 0.4em\relax
  ACM, 2002, pp. 347--356.

\bibitem{lim2014improving}
Y.-s. Lim, Y.-C. Chen, E.~M. Nahum, D.~Towsley, and R.~J. Gibbens, ``Improving
  energy efficiency of mptcp for mobile devices,'' \emph{arXiv preprint
  arXiv:1406.4463}, 2014.

\bibitem{lim2014green}
{Lim, Yeon-sup and Chen, Yung-Chih and Nahum, Erich M and Towsley, Don and
  Gibbens, Richard J}, ``How green is multipath {TCP} for mobile devices?'' in
  \emph{Proceedings of the 4th workshop on All things cellular: operations,
  applications, \& challenges}.\hskip 1em plus 0.5em minus 0.4em\relax ACM,
  2014, pp. 3--8.

\bibitem{limvalidation}
Y.-s. Lim, Y.-C. Chen, E.~M. Nahum, D.~Towsley, R.~Gibbens, and E.~Cecchet,
  ``Validation of energy aware {MPTCP} in the wild.''

\bibitem{nicutar2013evolving}
C.~Nicutar, C.~Paasch, M.~Bagnulo, and C.~Raiciu, ``Evolving the internet with
  connection acrobatics,'' in \emph{Proceedings of the 2013 workshop on Hot
  topics in middleboxes and network function virtualization}.\hskip 1em plus
  0.5em minus 0.4em\relax ACM, 2013, pp. 7--12.

\end{thebibliography}

\end{document}